\documentclass[aps,twocolumn]{revtex4-2}
\usepackage[bookmarksnumbered,bookmarksopen,colorlinks,citecolor=red,linkcolor=blue]{hyperref}
\usepackage{graphicx}
\usepackage{bm}
\usepackage{amsmath}
\usepackage{xcolor}
\usepackage{booktabs}
\usepackage{multirow}
\usepackage{diagbox}

\graphicspath{{./figs/}}

\begin{document}
\title{System size dependence of energy loss and correlations of heavy mesons at LHC energies }
\author{Jiaxing Zhao, Joerg Aichelin, Pol Bernard Gossiaux, and Klaus Werner}
\affiliation{SUBATECH, Nantes University, IMT Atlantique, IN2P3/CNRS, \\
4 rue Alfred Kastler, 44307 Nantes cedex 3, France}

\date{\today}

\begin{abstract} 
We study the system size dependence of heavy quark (HQ) observables at a center of mass energy of $\sqrt{s_{\rm NN}}=5.02$ TeV  to explore whether it can provide further constraints on the physical processes which are involved: energy loss of HQs in the quark gluon plasma (QGP), hadronization and hadronic rescattering. We use the EPOS4HQ approach to study p-p and 0-10\% central O-O, Ar-Ar, Kr-Kr and Pb-Pb reactions and investigate in detail the momentum change of heavy quarks from creation until their detection as part of a hadron as well as the enhancement of heavy baryon production at low $p_T$.  We investigate furthermore the origin and the system size dependence of the azimuthal correlations between the heavy quark $Q$ and the heavy antiquark $\bar Q$ and how one can bypass the problem that correlations are washed out due to the combinatorial background. We conclude that a systematic study of the system size dependence of the momentum loss allows to separate the momentum loss due to the passage through the QGP from the  momentum change due to hadronization and the $Q\bar Q$ correlations allow to gain inside into the different pQCD processes in which the heavy quarks are created. 
\end{abstract}


\maketitle

\section{Introduction}
\label{sec.intro}

Theory predicts that at high temperature and/or high density a plasma of   quarks and gluons (QGP) is formed in which partons are not confined anymore. One of the main goals of heavy-ion reactions at ultra-relativistc energies is to study this new form of deconfined matter. In the last years ample evidence has been gathered that this state exists. The observation of strangeness enhancement~\cite{STAR:2011fbd}, quark scaling of the elliptic flow~\cite{PHENIX:2006dpn}, jet quenching~\cite{CMS:2016xef}, and quarkonium suppression~\cite{Matsui:1986dk} points in this direction.  It is the task of future experiments and the challenge for theory to study the properties of the QGP, not only at the moment when it hadronizes into mesons and baryons but also before, during the expansion of the QGP when the temperature is considerably higher than the hadronization temperature of about $T_c \approx 160 \rm  \ MeV$.
Because the multiplicity of mesons, which contain only light valence quarks, is well described by a statistical model with $T=T_c$ and the spectra of light hadrons are perturbed by hadronic rescattering, the efforts concentrated in the last year on the study of penetrating probes, which include jets and heavy flavour hadrons.
Heavy flavour quarks have the advantage that they are created only in the initial hard collisions between projectile and target nucleons because their mass $m_Q$ is large compared to QGP temperatures and that, due to $m_Q >\Lambda_{QCD}$, their production can be described by perturbative QCD (pQCD). 

In the last years many theoretical approaches have been advanced to describe the passage of heavy quarks through the medium, the subsequent hadronization and hadronic rescattering. They include PHSD \cite{Song:2015sfa,Song:2015ykw}, in which the heavy quark physics is embedded in an approach, which described the light hadrons as well. It is based on the dynamical quasi particle model (DQPM), which respects the equation of state of strongly interacting matter. The Catania model \cite{Plumari:2017ntm,Scardina:2017ipo,Minissale:2020bif} is also based on
a DQPM approach and describes the expanding medium by a Boltzmann equation. The LBT model \cite{Cao:2016gvr} solves as well a Boltzmann equation for the heavy quarks including elastic and inelastic collisions but the medium is described by viscous hydrodynamics. Other models like TAMU\cite{He:2019vgs,He:2019tik}, Duke\cite{Cao:2015hia}, and Torino \cite{Beraudo:2022dpz,Beraudo:2023nlq} replace the Boltzmann equation by a Langevin equation to describe the dynamics of heavy quarks and model the expanding QGP by ideal or viscous hydrodynamics, fitted to different key observables of light hadrons. 

Recently we advanced the EPOS4HQ model, which is based on the EPOS4 approach, \cite{Werner:2023fne,Werner:2023jps,Werner:2023mod,Werner:2023zvo}, to describe the physics of heavy quarks.  EPOS4 has been very successful in describing a multitude of light hadron observables from RHIC to LHC energies \cite{Werner:2023mod}. 
The agreement between EPOS4 and the experimental observables indicates that the time evolution of the QGP, including hadronization and hadronic rescattering, is well reproduced.  EPOS4 creates heavy quarks as well but they do not interact with the QGP. EPOS4HQ embeds the heavy quark physics in the EPOS4 approach. There the heavy quarks,  created in the EPOS4 interaction points,  interact with the QGP partons by elastic and inelastic collisions, described by pQCD elastic and inelastic cross sections. Heavy quarks, which have passed a QGP,  hadronize either by coalescence or by fragmentation. The relative weight of both processes depends on the transverse momentum, $p_T$, of the heavy quark.  A detailed description of EPOS4HQ, including the scattering with the plasma constituents and hadronization,  can be found in \cite{Zhao:2023nrz}. There is also shown that EPOS4HQ describes all the presently available data quite well.
 
In \cite{Zhao:2023ucp} the EPOS4HQ approach has been applied to pp collisions and it has been demonstrated that it  describes equally well the heavy hadron production in pp collision, assuming that in  pp collisions  as well as, in heavy-ion collisions the same criterion for the formation of a QGP applies, see Ref. \cite{Werner:2023jps}. 

The agreement of the EPOS4HQ approach with p-p as well as with Pb-Pb data motivates to study systematically the system size dependence of several key observables. We start out with a short overview over the heavy quark production processes (section~\ref{sec.2}) and of the QGP in EPOS4 (section~\ref{sec.3}). The momentum difference between the $c$ and $\bar c$-quark at creation and between the finally observed $D$-mesons as well as the momentum loss of HQ in a QGP is shown in section~\ref{sec.4}. This is followed by a study of  the system size dependence of the heavy baryon enhancement (section~\ref{sec.5}) as well as of the change of the azimuthal correlation between a $D$ and a $\bar D$ meson, created in the same elementary collision, due to the QGP (section~\ref{sec.6}). Finally, in section~\ref{sec.7}, we present our conclusions.

The calculations are based on version EPOS4.0.1.s9.

\section{heavy quark production and evolution in EPOS4}
\label{sec.2}
In  EPOS4HQ  the heavy quarks are produced together with light quarks and gluons. Energy and momentum is conserved at each vertex. The interaction vertices are given by EPOS4. Therefore the kinetic variables as well as the initial heavy quark coordinates are identical in EPOS4 and EPOS4HQ.  The difference is, as said, that in EPOS4HQ we follow the heavy quarks through the QGP where they may interact with the QGP partons and that heavy quarks, which pass a QGP, may hadronize not only by fragmentation but also by coalescence. 
\begin{figure}[!htb]
\includegraphics[width=0.5\textwidth]{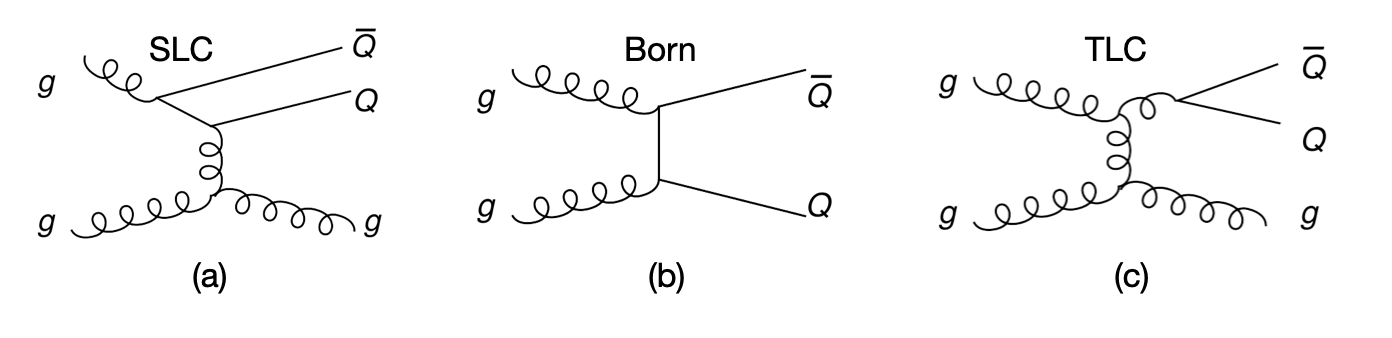}
\caption{Different possibilities to create heavy flavor: (a) flavor excitation (space-like cascade (SLC)), (b) flavor creation (Born process), (c) gluon splitting (time-like cascade (TLC)).}
\label{fig.hardprocess1}
\end{figure}

In EPOS4 $Q\bar Q$ pairs can be produced in different processes, which are shown in  Fig.~\ref{fig.hardprocess1}. Before the partons from projectile and target interact in a hard process they develop a space like DGLAP cascade (SLC), displayed in the left panel of Fig.~\ref{fig.hardprocess1}. In the DGLAP process gluons can be emitted, which can split into a heavy quark-antiquark pair. This is a next to leading order process. In the hard process,  Fig.~\ref{fig.hardprocess1}, middle, collisions of gluons or light quark-antiquark pairs can create a $Q\bar Q$ pair as well. This is the leading order (Born) process. The gluons, created in the hard process, are time like (TLC) and may therefore, in a next to leading order process, disintegrate into a $Q\bar Q$ pair. This process is shown in the right panel of    Fig.~\ref{fig.hardprocess1}.  As we will see later,  the correlations between the $Q$ and the $\bar Q$  are rather different in the three processes.

In EPOS4HQ heavy quarks interact with the partons of the QGP, formed by the light core partons and gluons. We include  both, elastic~\cite{Gossiaux:2008jv} and radiative~\cite{Aichelin:2013mra} collisions.  To calculate the energy  exchange of the heavy quark with the medium we first calculate the interaction rate for elastic and inelastic collisions. If an interaction takes place we select randomly the momentum of the scattering partner, a light quark or a gluon, from their corresponding thermal distribution. The thermal distribution is determined by the local temperature and the local mean parton velocity, at the position where the heavy quark is localized. The parton momentum is then boosted into the computational frame, the projectile/target center of mass system in which the heavy quark momentum is defined. The scattering cross sections of the heavy quark with gluons and light quarks are calculated by pQCD matrix elements with a running coupling constant. The kinematic variables of the final state particles are drawn randomly from the cross section integrated over energy and momentum conserving delta functions.

When the QGP has expanded to the critical value of the local temperature or energy density  ($T_{\rm FO}=167\ \rm MeV$ or $\epsilon_{\rm FO}=0.57\ \rm GeV/fm^3$), the system hadronizes and the heavy quarks, which are localized in the QGP, are converted into a heavy flavor (HF) hadron. In the EPOS4HQ framework, there are two ways in which the  heavy quark can hadronize, either by fragmentation or by coalescence. The momentum distribution of a HF hadron, which is produced via the coalescence process, can be calculated via,
\begin{eqnarray}
{dN\over d^3{\bf P}}&=&\sum g_H\int \prod_{i=1}^k{d^3p_i\over (2\pi)^3E_i} f_i({\bf p}_i)\nonumber\\
&\times&W_H({\bf p}_1,..,{\bf p}_i)\, \delta^{(3)}\left({\bf P}-\sum_{i=1}^k{\bf p}_i\right),
\label{eq.coal}
\end{eqnarray}   
where $g_H$ is the degeneracy factor of color and spin. $k$ equals 2(3) for mesons (baryons). ${\bf P}$ and ${\bf p}_i$ are the momenta of heavy flavor hadrons and the quark momenta, respectively. The delta function conserves the momentum.  
$f_1({\bf p}_1)=(2\pi)^3\delta^{(3)}\left({\bf p}_c-{\bf p}_1\right)$ is the normalized momentum space distribution of the heavy quark and  $f_i({\bf p}_i)$ for $i>1$ is the momentum space distribution of the light quarks at the hadronization point.  $W_H({\bf p}_1,..,{\bf p}_i)$ is the momentum distribution of the quarks bound in a heavy hadron $H$ in momentum space, corresponding to the Wigner density distribution after integration over coordinate space. It can be constructed from the hadron wave function. The hadron wave function can be approximated by a three-dimensional harmonic oscillator state with the same root mean square radius. For the ground state of charmed mesons, the Wigner density in the center-of-mass (CM) frame can be expressed as,
\begin{eqnarray}
W(p_r)&=&{(2\sqrt{\pi}\sigma)^3}e^{-\sigma^2{p_r}^2},
\end{eqnarray} 
where $p_r ={|E_2{\bf p}_1-E_1{\bf p}_2|/( E_1+E_2)}$ is the relative momentum between the  two constituent quarks in their CM frame. It is normalized, $\int W(p_r)d^3{\bf p}_r/(2\pi)^3=1$. 
$E_1 ({\bf p}_1)$ and $E_2({\bf p}_2)$ are the energies and momenta of the quark or antiquark in the CM frame, respectively. The width $\sigma$ in the Wigner density is controlled by the root-mean-radius. 
If heavy quarks hadronize by fragmentation we apply  the HQET-based fragmentation function~\cite{Braaten:1994bz,Cacciari:2005rk}.
The probability for coalescence, $P_{coal}$, and of fragmentation, $1-P_{coal}$, depends on the momentum  of the heavy quark. It is displayed in \cite{Zhao:2023nrz}.
Fragmentation and coalescence act differently: Whereas the momentum of the $D$-meson formed by coalescence is usually larger than that of the $c$-quark due to the momentum of the light quark, which  is picked up, the momentum of the $D$-meson formed by fragmentation is smaller. 
\begin{figure*}[!htb]
\includegraphics[width=1.0\textwidth]{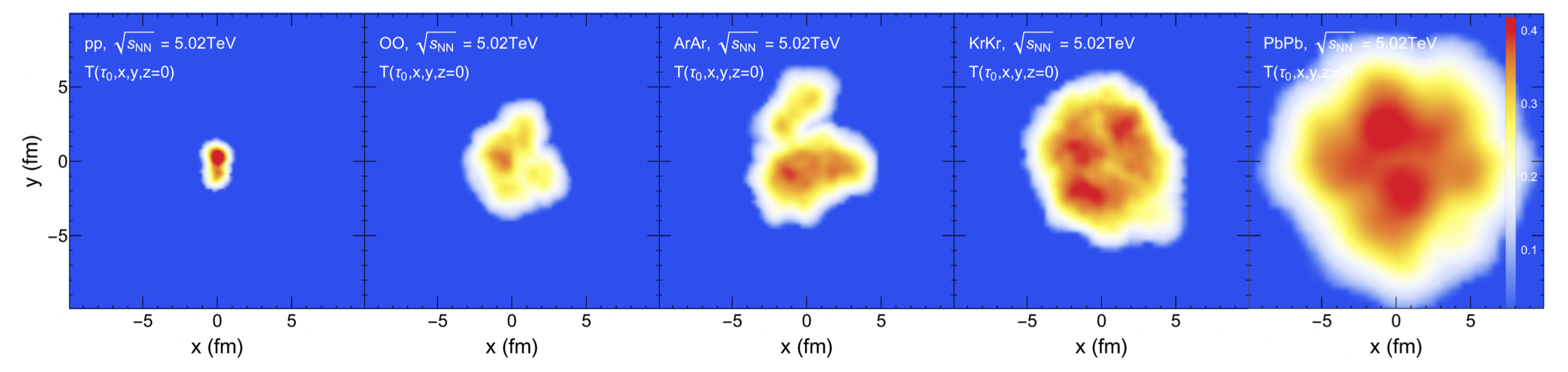}
\caption{Temperature profile of one event in the transverse plane at the initial time $\tau_0$ in p-p, O-O, Ar-Ar, Kr-Kr, and Pb-Pb collisions at $\sqrt{s_{\rm NN}}=5.02$ TeV and at $z=0$.}
\label{fig.Txy}
\end{figure*}

\section{Hot QCD medium in EPOS4}
\label{sec.3}
EPOS4 and EPOS4HQ use event by event hydrodynamics. This means that in every simulation the interaction points and the 
parton productions are different. The parton energy distribution is the initial condition for the hydrodynamic evolution
of QGP. 
Typical temperature profiles of a single event in the plane perpendicular to the beam direction for $z=0$ at the time $\tau_0$  ($\tau_0$=0.4, 1.024, 1.078, 1.167, 1.457 fm/c for p-p, 0-10\% O-O, Ar-Ar, Kr-Kr, and Pb-Pb collisions)  are shown in Fig.~\ref{fig.Txy} for central O-O, Ar-Ar, Kr-Kr, and Pb-Pb events.  $x$ is the direction of the impact parameter.  We observe strong fluctuations of the temperature whose scale is given in GeV.  Large areas are above the critical temperature of 0.167 GeV above which a QGP exists. So almost all heavy quarks,  in O-O as well as in Pb-Pb collisions, pass a QGP.  The hydrodynamic evolution of the QGP stops when the critical energy density of $\epsilon_{\rm FO}$ = 0.57 GeV/fm$^3$ is reached. Then the system hadronizes, as described in \cite{Werner:2023jps}.

\section{heavy quark momentum loss}
\label{sec.4}
\begin{figure*}[!htb]
\includegraphics[width=1.0\textwidth]{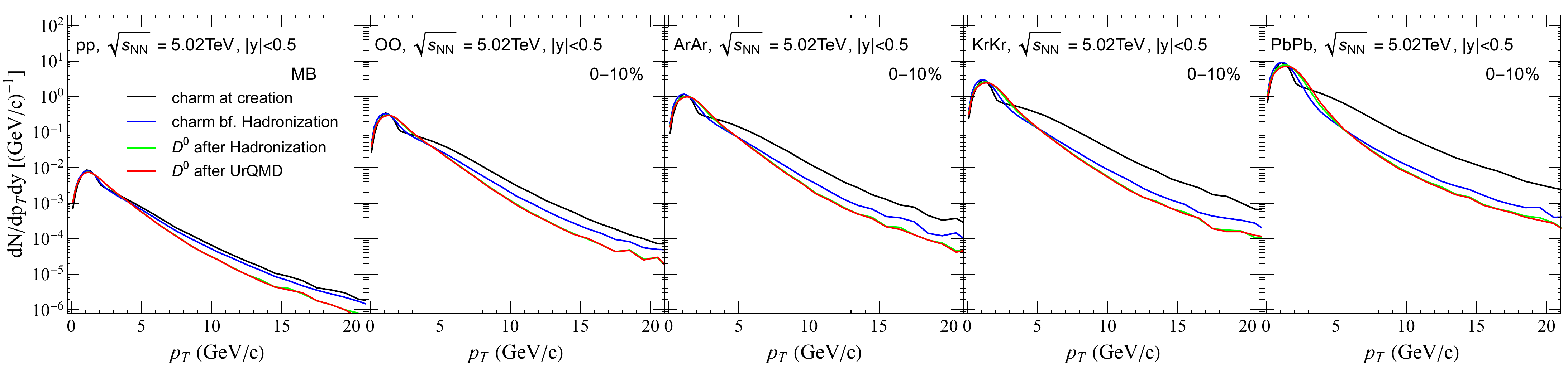}
\caption{$p_T$ spectra of charm quarks at creation, before hadronization as well as of $D^0$ after hadronization and UrQMD in p-p, O-O, Ar-Ar, Kr-Kr, and Pb-Pb collisions at $\sqrt{s_{\rm NN}}=5.02$ TeV and central rapidity $|y|<0.5$. Here only charm quarks are included, which are finally part of $D^0$ mesons.}
\label{fig:spectra}
\end{figure*}
\begin{figure}[!htb]
\includegraphics[width=0.25\textwidth]{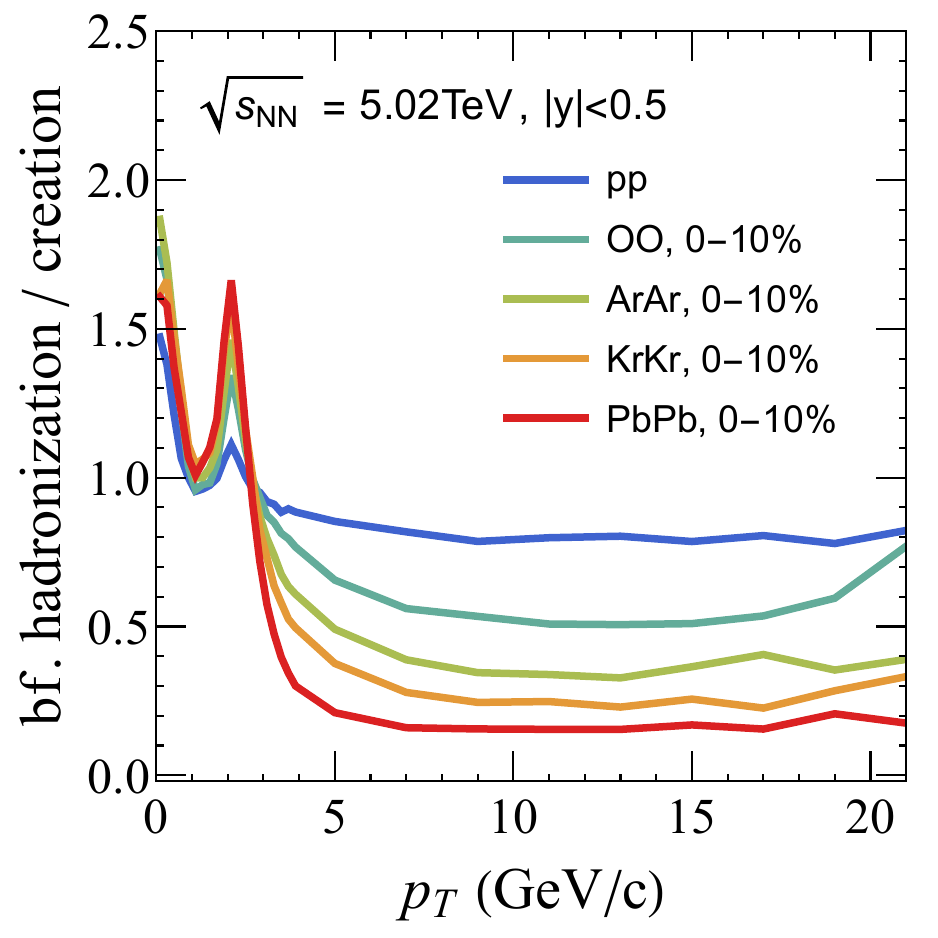}\includegraphics[width=0.25\textwidth]{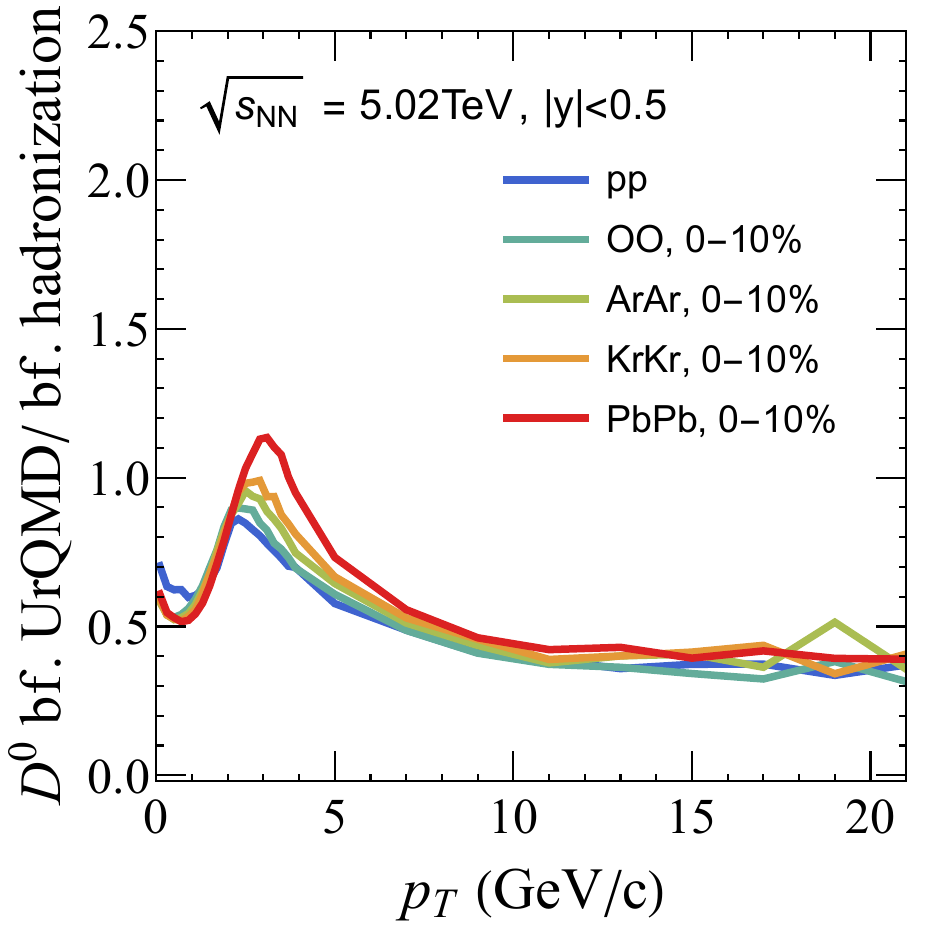}\\
\includegraphics[width=0.25\textwidth]{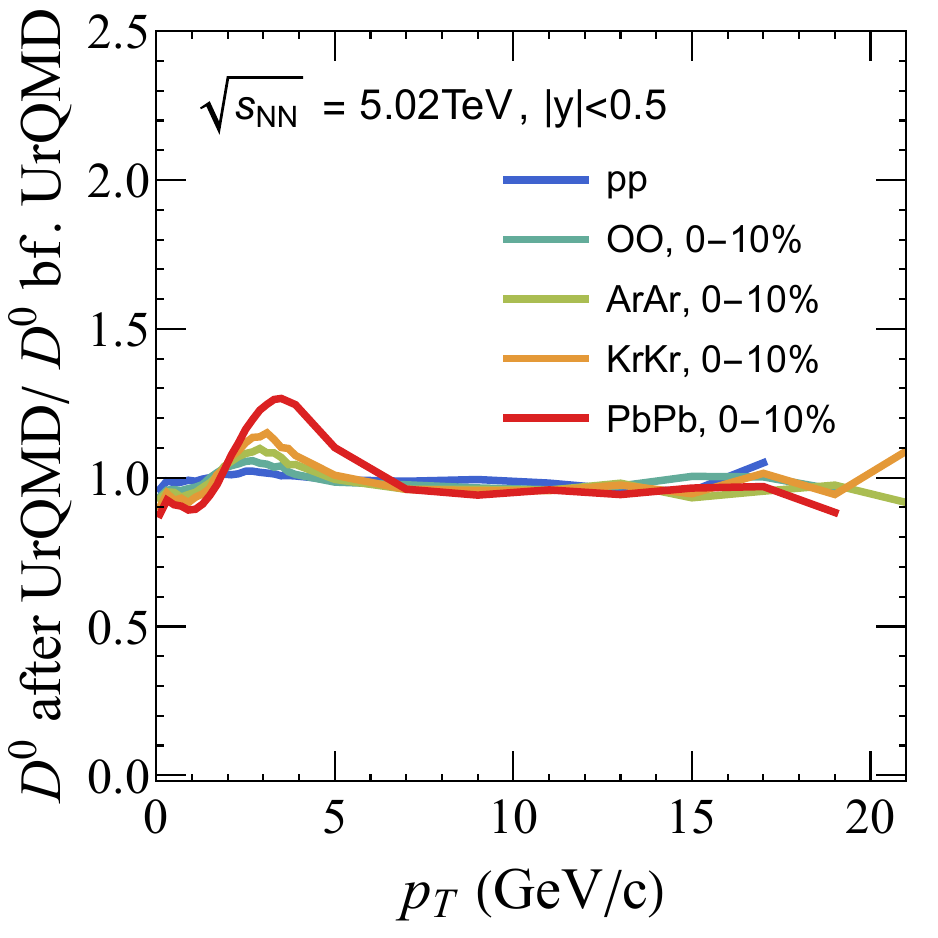}
\caption{Ratio of the momentum spectrum of charm quarks before hadronization and that of charm quarks at creation(top left)  as well as that of  final $D^0$ meson and charm quarks before hadronization (top right) in p-p, O-O, Ar-Ar, Kr-Kr, and Pb-Pb collisions at $\sqrt{s_{\rm NN}}=5.02$  TeV and central rapidity $|y|<0.5$.  The bottom figure displays the ratio of the spectra of $D$-mesons after and before hadronic rescattering.} 
\label{fig:ratio}
\end{figure}
On the way to the detector heavy quarks, which are later converted into heavy mesons, change their momentum by three processes: a) by interaction with the QGP partons, b) by hadronization, when the heavy quark becomes a heavy hadron and c) by hadronic rescattering. 

In Fig.~\ref{fig:spectra}  we display the transverse momentum distribution of heavy quarks and heavy mesons created at midrapidity ($|y|<0.5$) in minimum bias (MB) for p-p and central collisions for different symmetric heavy-ion collisions. We select here only those heavy quarks which hadronize to $D^0$-mesons. Therefore the norm of the spectra for $c$-quarks and $D$-mesons
is identical. The black line shows the distribution of the charm quarks when they are created in the pQCD processes, the blue line that at the moment before they hadronize into $D^0$-mesons. The red and green lines show the transverse momentum spectrum of the $D$-mesons immediately after hadronization and after hadronic rescattering, described by UrQMD, respectively. 

We see, first of all, that the red and green lines are almost identical, so the hadronic rescattering does not modify the spectra considerably. We observe furthermore that the difference between the black and the blue lines increases with the size of the colliding system, as we expect, because the spatial extension of the QGP increases with the system size and we expect a larger momentum loss if system gets larger. It is remarkable that already in p-p collisions the black and blue line differ, indicating the formation of a QGP, as we have pointed out recently \cite{Zhao:2023ucp}. Hadronization changes the spectra in a non negligible way. For $c$-quarks with a  $p_T > 5$ GeV hadronization by fragmentation is the dominant process and so we expect that the $p_T$ of the $D$-mesons is smaller that that of the c-quark before hadronization. This shift is seen in the spectra. At low $p_T$ coalescence dominates and therefore we expect a momentum gain in the hadronization process, as also seen in the spectra.

To elucidate the physics from a different point of view we display in Fig.~\ref{fig:ratio} ratios of the $p_T$ spectra for different projectile-target combinations. Top left we display the ratio of the $c$-quark spectra just before hadronization and at creation. Top right shows the ratio of the $D$-meson $p_T$ distribution after hadronization and the $p_T$ distribution of the charm quark before hadronization, hence the change of the spectrum due to hadronization. The bottom figure shows the ratio of the $D$-meson spectra after and before UrQMD and consequently the influence of the hadronic rescattering. 

For $p_T>5$ GeV the ratio before hadronization/at creation depends strongly on the system size. For Pb-Pb the spectrum before hadronization is suppressed by more than a factor of  5 but even for a system as small as O-O the suppression  is close to a factor of 2, so clearly visible.
At low $p_T$ the heavy quark rescattering leads to a partial thermalization of the heavy quarks with the plasma partons. This enhances the hadronization $p_T$ spectra around the thermal equilibrium value of $p_T$. The ratio of the $p_T$ spectrum of the $D$-meson immediately after hadronization and of the $c$-quark before hadronization, displayed on the right hand side, shows that the $D$-meson spectrum is suppressed as compared to that of the $c$-quarks. For $p_T > 5$ GeV this suppression of the order of 2 is almost independent of the size of the colliding system.  
This confirms that hadronization is a local process, which happens when the expanding QGP has reached the critical temperature and which is therefore almost independent of the system size.  This shift is a superposition  between the momentum increase due to hadronization by coalescence and the momentum loss due to hadronization by fragmentation. Fig.~\ref{fig:ratio}  bottom, shows the ratio of the $p_T$ spectrum after and before hadronic rescattering, which is close to one for 
$p_T > $ 5 GeV.
\begin{figure}[!htb]
\includegraphics[width=0.25\textwidth]{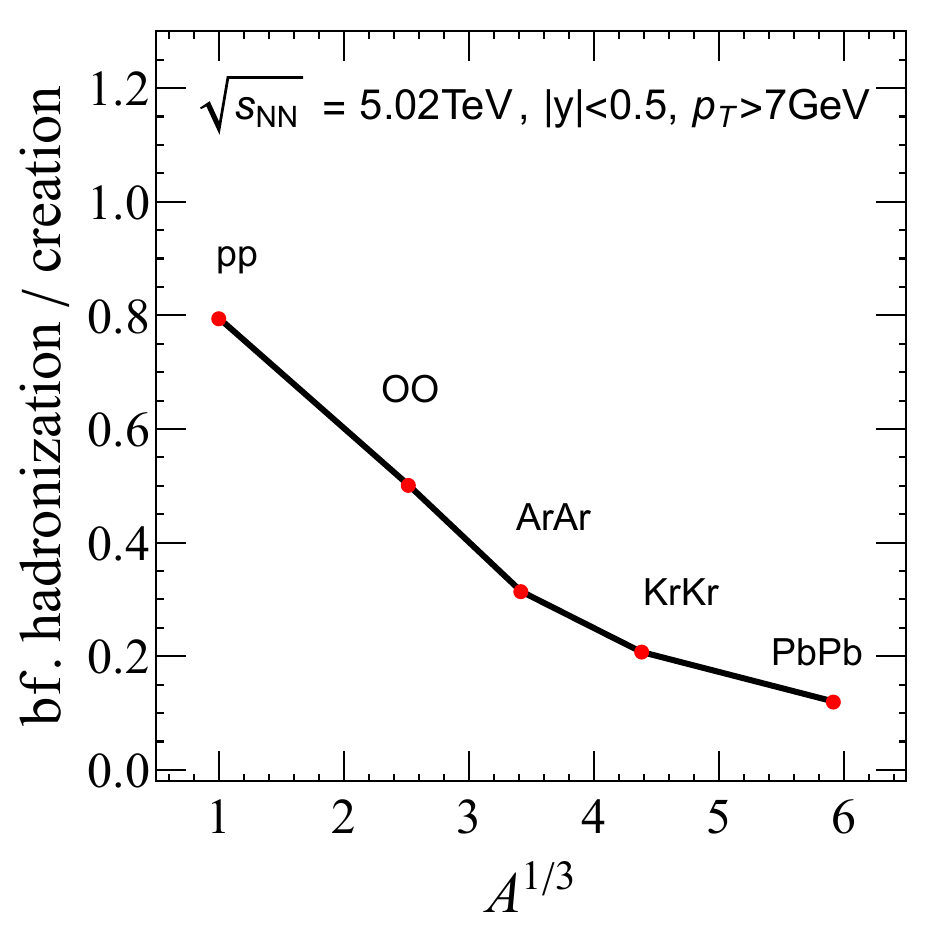}\includegraphics[width=0.25\textwidth]{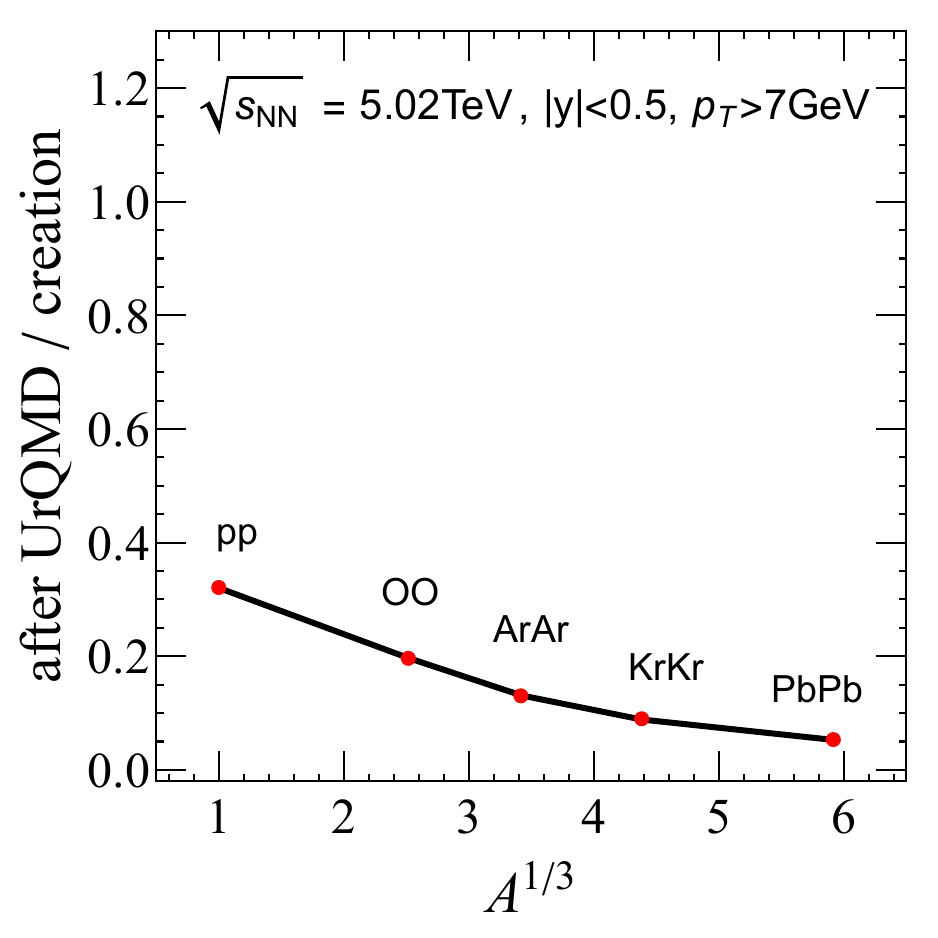}\\
\caption{System size dependence of the ratio of $p_T$ integrated transverse momentum spectra for  $p_T>7\;\rm GeV/c$. On the left we display the ratio before hadronization/at creation, on the right the ratio after UrQMD/at creation for  $\sqrt{s_{\rm NN}}=5.02$ TeV and at central rapidity $|y|<0.5$. }
\label{fig.adep}
\end{figure}

In Fig.~\ref{fig.adep} we compare  integrated $p_T$ spectra (for $p_T > 7$ GeV)  as a function of $A^{1/3}$ to quantify more their system size dependence. On the left hand side we present the ratio before hadronization/at creation, on the right hand side before UrQMD/at creation. As shown, the final spectrum equals the spectrum before UrQMD. 
We see again from the left figure that larger systems shift the high momentum particles more to lower $p_T$ values than smaller systems. In Pb-Pb 90\% of the heavy quarks with an initial  $p_T > 7 $ GeV are shifted to momenta below 7 GeV. In p-p collisions this is only the case for 20\%.  For smaller systems the fraction of $c$-quarks shifted to lower $p_T$ is roughly proportional to $A^{1/3}$, and so proportional to the average length, which the heavy quarks have to travel in the QGP.  If we compare the ratio after UrQMD/at creation on the left hand side we find a curve which is characteristic for the relative fraction of  the momentum loss 
due to collisions in the QGP and due to hadronization.  This curve can be normalized 
to the pp values and compared to the experimental finding.  Therefore, a systematic study of the system size dependence offers the possibility to separate experimentally the momentum loss of heavy quarks in the QGP (which increases with the system size) and that due to hadronization (which is system size independent and dominates at small systems) just by comparing the integrated  $p_T$ spectra of $D$ mesons. This would also allow to calibrate better the transport models which do not agree on the relative fraction of both momentum losses \cite{Xu:2018gux,Zhao:2023nrz}.

The microscopic description of the time evolution of heavy quarks in EPOS4HQ allows to follow the heavy quarks individually until their hadronization. We can therefore calculate the momentum transfer, which a heavy quark suffers while it is moving in the QGP, as well as the momentum difference between the $c$-quark before hadronization and the $D$-meson after hadronization.
\begin{figure}[!htb]
\includegraphics[width=0.25\textwidth]{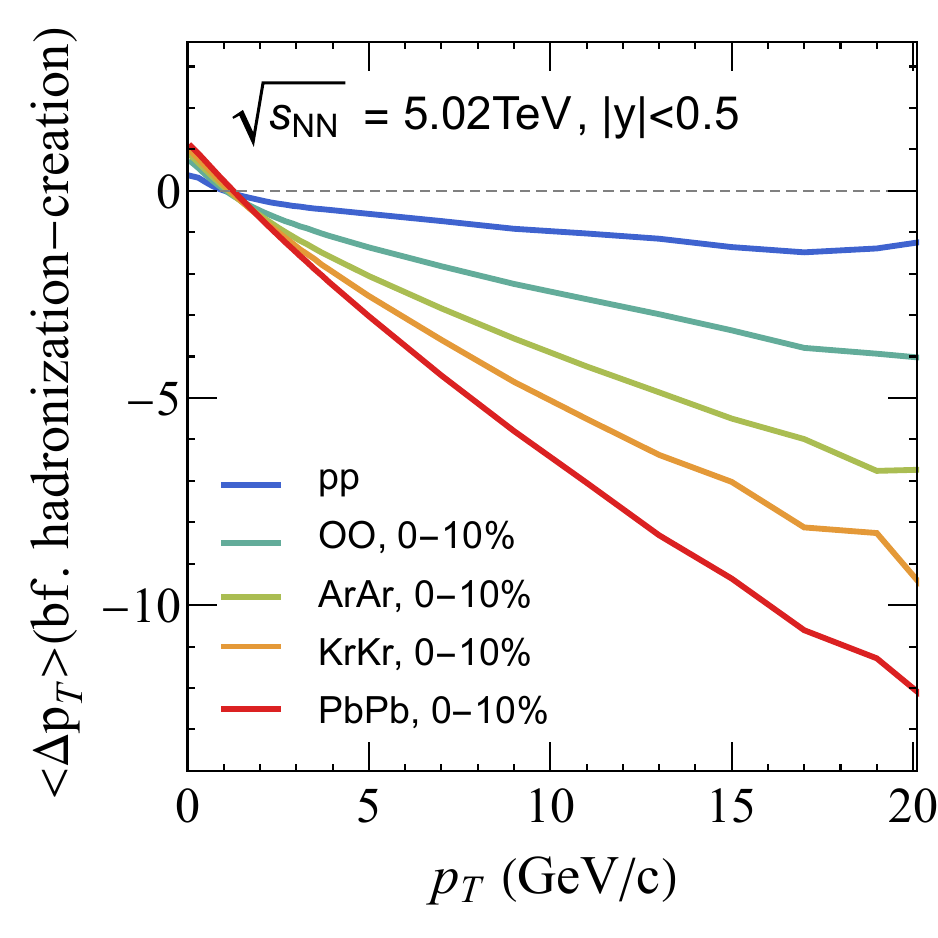}\includegraphics[width=0.25\textwidth]{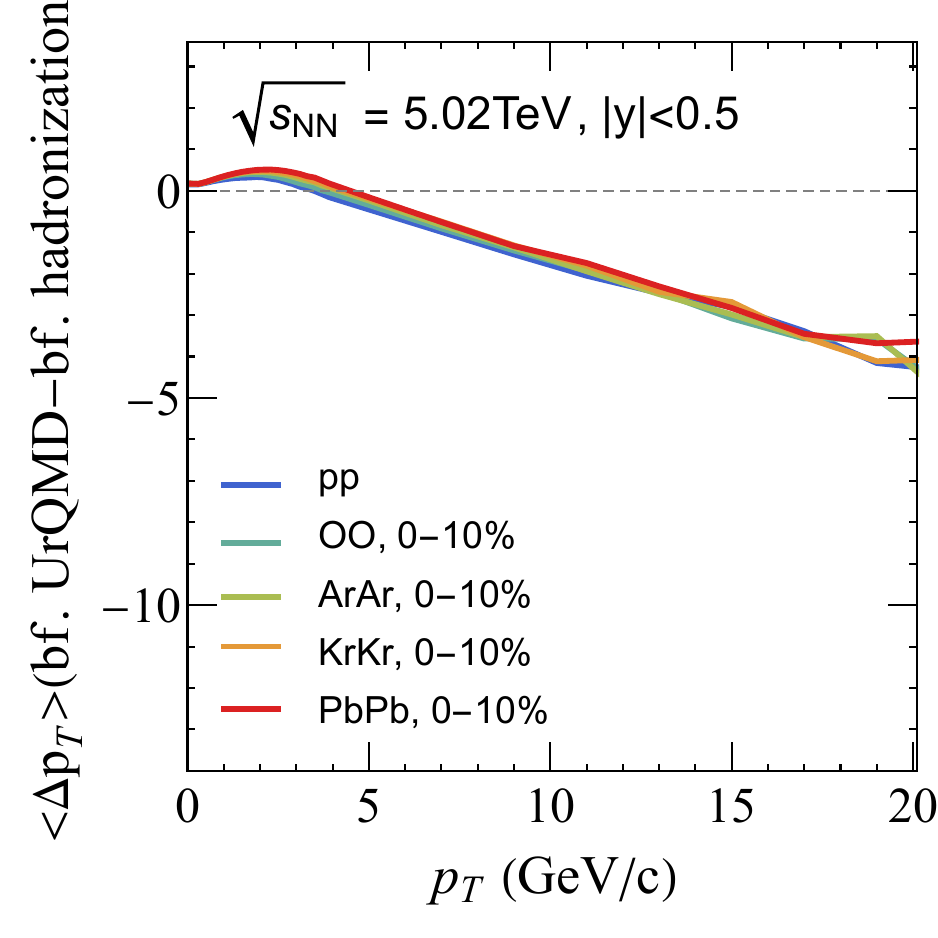}\\
\includegraphics[width=0.25\textwidth]{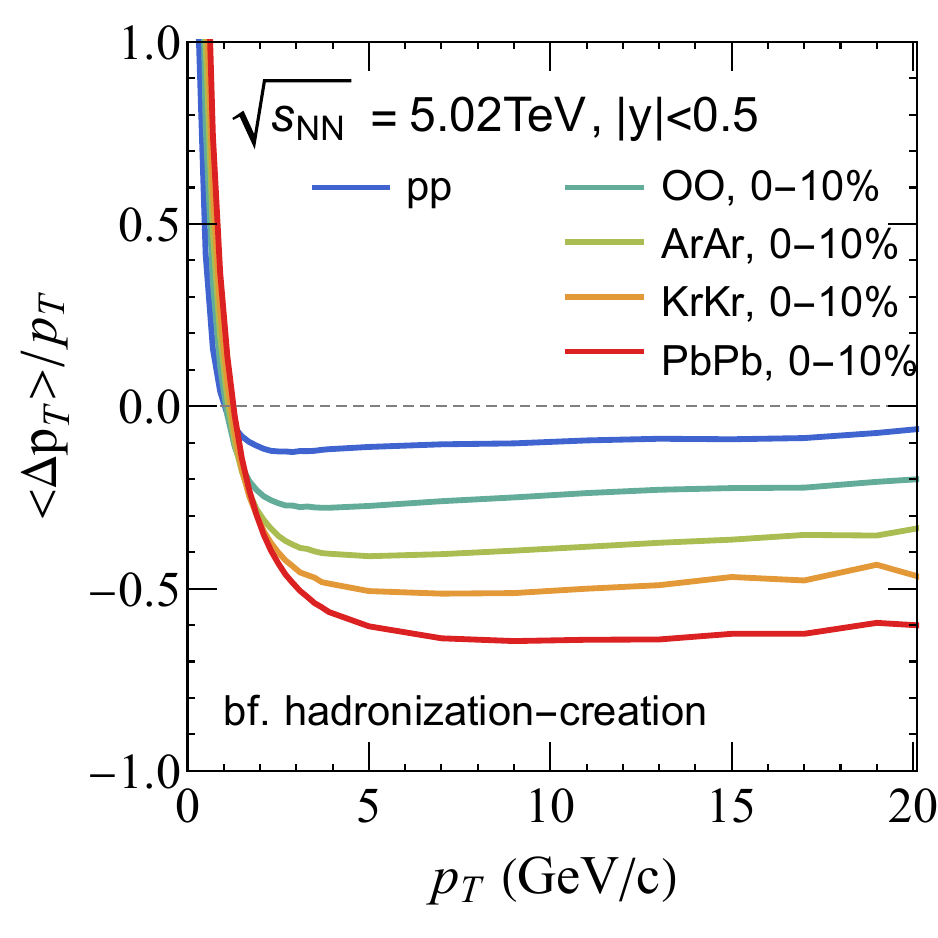}\includegraphics[width=0.25\textwidth]{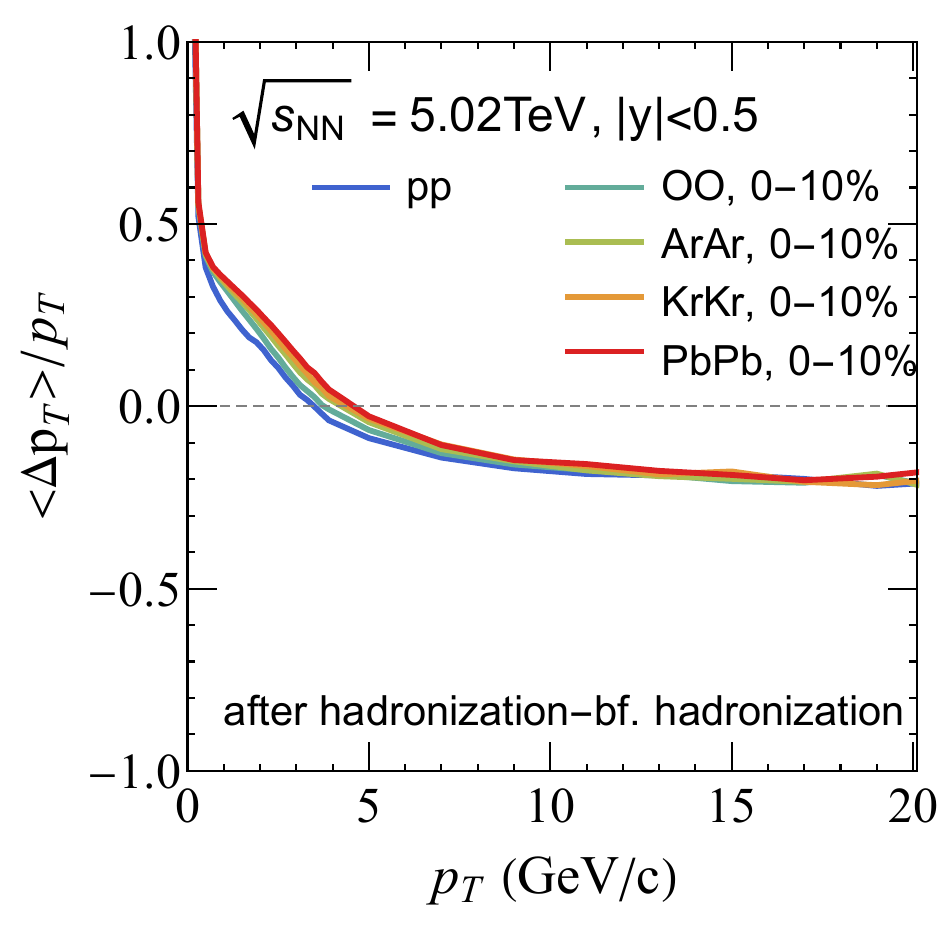}\\
\caption{Transverse momentum change in the hot medium (left) and during hadronization (right) in p-p, O-O, Ar-Ar, Kr-Kr, and Pb-Pb collisions at $\sqrt{s_{\rm NN}}=5.02$ TeV and central rapidity $|y|<0.5$. $\Delta p_T$ is in GeV.}
\label{fig:shift}
\end{figure}
For $\sqrt{s_{\rm NN}} =$ 200 GeV Au-Au collisions, this momentum loss has already been calculated using another hydrodynamical model \cite{Gossiaux:2009mk}. There a momentum shift of approx 7 GeV was observed for an initial $c$-quark momentum of 20 GeV. 

In Fig.~\ref{fig:shift} we present the results for $\Delta p_T$ (top row), as well as those
for $\Delta p_T/p_T$ (bottom row). On the left hand side we see, for large $p_T$, an increase of the momentum loss in the QGP  with $p_T$. Because for all systems  $\Delta p_T/p_T$ is almost constant at high $p_T$, the momentum change of a heavy quark in a Pb-Pb collisions with $p_T$= 20 GeV is almost a factor of four larger than for a quark at 5 GeV. This is a consequence of the energy dependence of the cross sections and the kinematic of the reaction. Although for all systems the momentum shift as a function of $p_T$ is almost linear, the slope is different for the different systems due to different lengths, which the heavy quarks travel in the QGP. At low $p_T$, where the heavy quarks are slow and stay for a long time in contact with the QGP particles, we see that the $c$-quarks are in the process to come to an equilibrium with the QGP, as already observed in \cite{Nahrgang:2013saa}. Therefore the heavy quarks may gain momentum.   

Also the momentum shift due to hadronization, displayed at the right hand side, depends on $p_T$. Here we observe roughly a factor of two between $p_T$= 10 GeV and $p_T$= 20 GeV: The fragmentation function scales with the momentum fraction of the $D$-meson with respect to the momentum of the $c$-quark. Hence the momentum shift of a $D$-meson produced by a $p_T$ = 10 GeV $c$-quark should be roughly half of that produced by a $p_T$ = 20 GeV $c$-quark, what is indeed observed. The momentum shift due to hadronization is system size independent due to the local character of hadronization. At low $p_T$ the momentum change is positive due to the coalescence process.

\begin{figure}[!htb]
\includegraphics[width=0.35\textwidth]{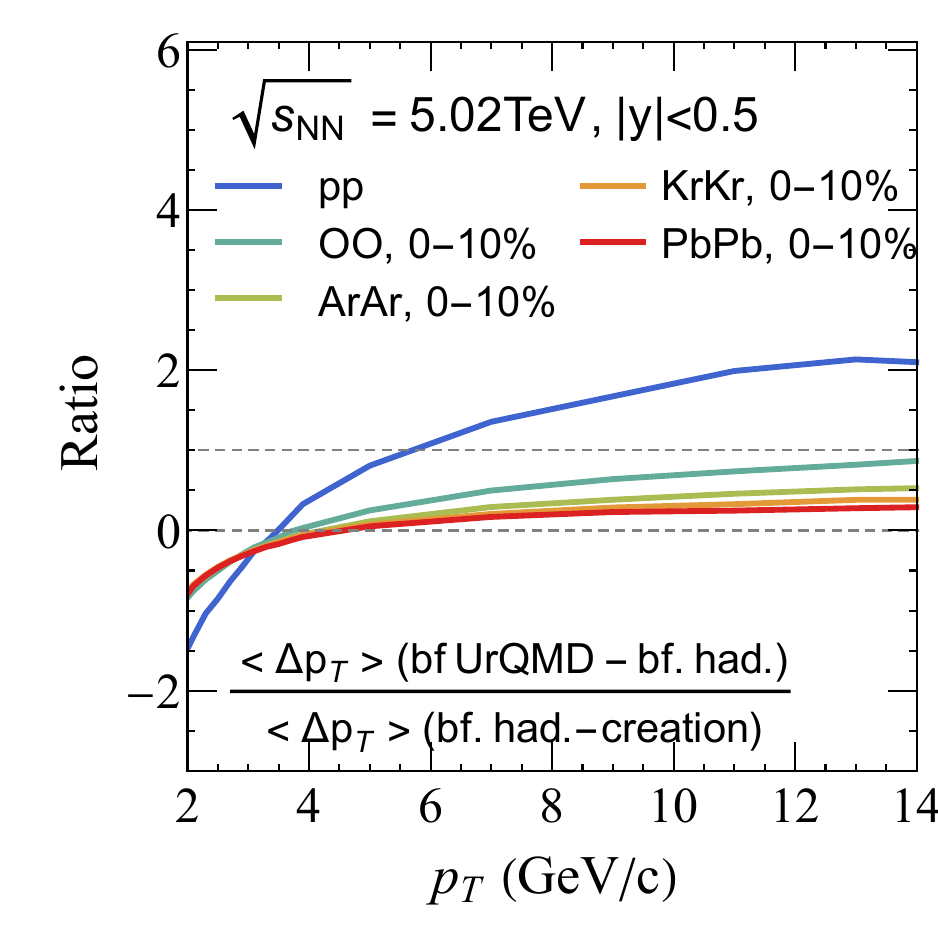}
\caption{Ratio of the transverse momentum shift in the QGP and due to hadronization for  p-p, O-O, Ar-Ar, Kr-Kr, and Pb-Pb collisions at $\sqrt{s_{\rm NN}}=5.02$ TeV and at central rapidity $|y|<0.5$. The two horizontal lines serve to guide the eye.}
\label{fig:shift2}
\end{figure}

To explore the relative importance of the physical processes to which the finally observed $D$-mesons (here are all charmed hadrons before UrQMD) are sensitive to, we divide both ratios, the ratio of the momentum shift of the $c$-quark due to interactions with the QGP and the momentum shift caused by hadronization. The result is displayed in Fig.~\ref{fig:shift2}. 
Let us start with the high $p_T$ part. Since both, momentum shift in the plasma and momentum shift due to hadronization, increase with $p_T$, the ratio is rather flat. For light projectile-target combination the momentum shift due to hadronization is larger than that due to interactions with the QGP. The final $D$-meson $p_T$ spectrum is there more sensitive to the fragmentation process (and its correct description in the medium) then to the energy loss of the $c$-quark in the QGP and is therefore little informative about the $c$-quark interactions with the QGP. For larger systems this ratio may decrease  by a factor of three and therefore the final spectra become more sensitive to the $c$-quark interactions with the QGP.  
\section{Multiplicity of Heavy Baryons}
\label{sec.5}
 
\begin{figure*}[!htb]
\includegraphics[width=1.0\textwidth]{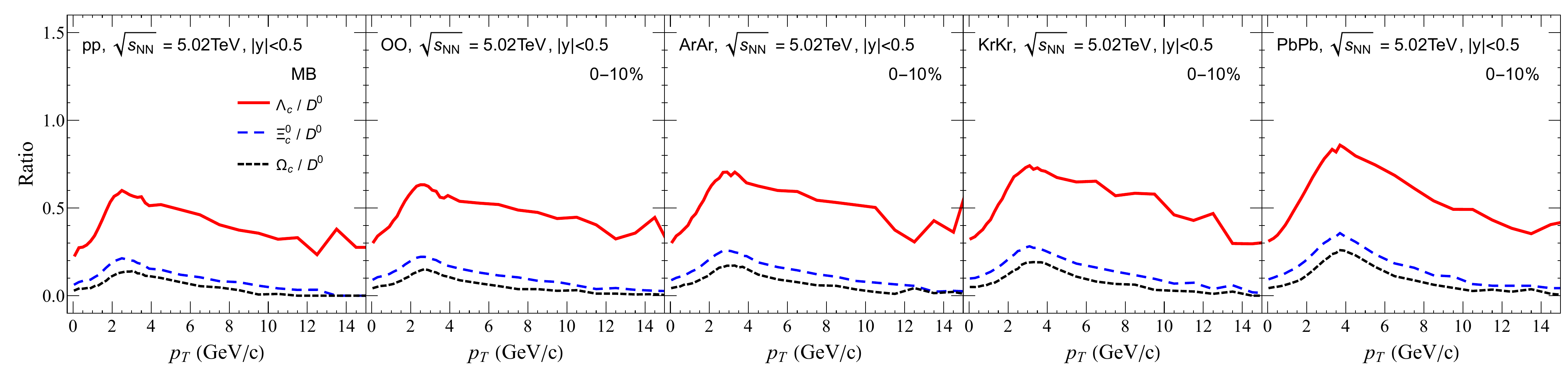}
\caption{Yield ratio of $\Lambda_c/D^0$, $\Xi_c^0/D^0$, and $\Omega_c/D^0$ in p-p, O-O, Ar-Ar, Kr-Kr, and Pb-Pb collisions at $\sqrt{s_{\rm NN}}=5.02$ TeV and central rapidity $|y|<0.5$. }
\label{fig.yieldratio}
\end{figure*}
The enhanced production of heavy baryons at low $p_T$ as compared to the value expected
from $e^+e^-$ fragmentation functions was one of the surprising results in p-p collisions. Such an enhancement has also been observed in Pb-Pb and there it is attributed the the hadronization of heavy quarks at the end of the expansion phase of the QGP, where heavy quarks may form heavy baryons by coalescence. A smooth system size dependence of this enhancement may therefore verify the conjecture that also in very small systems like p-p a QGP is formed, where heavy baryons can be produced by coalescence. This conjecture is based on the successful description of different observables in p-p collision in the EPOS4HQ approach~\cite{Zhao:2023ucp}.

The system size dependence of the enhancement of heavy baryons at low $p_T$ is shown in Fig. 
\ref{fig.yieldratio}, which displays from left to right the ratio of the $p_T$ spectra of
$\Lambda_c/D^0, \Xi_c/D^0$ and $\Omega_c/D^0$ for p-p, O-O, Ar-Ar, Kr-Kr and Pb-Pb collisions.  We observe
that this enhancement is rather similar for all systems even if a detailed analysis shows that the maximum of the enhancement shifts from $p_T\approx 3$ GeV for p-p to $p_T\approx $4 GeV for Pb-Pb.  Also the enhancement at the maximum increases by 40\% from p-p to Pb-Pb.  If this smooth system size dependence, and explicitly the small difference between p-p and O-O, is verified by experiments it is hard to imagine that a p-p collision proceeds differently than a heavy ion collision.

\section{Correlations}
\label{sec.6}
Another source of information, not only on the interaction of $c$-quarks with the QGP but also on the $c$-quark production mechanism itself, are correlations among heavy hadrons. They have been measured in p-p collisions at forward rapidity by the LHCb collaboration~\cite{LHCb:2012aiv} and been successfully described by EPOS4HQ \cite{Zhao:2023ucp,Werner:2023zvo}. 
\begin{figure}[!htb]
\includegraphics[width=0.45\textwidth]{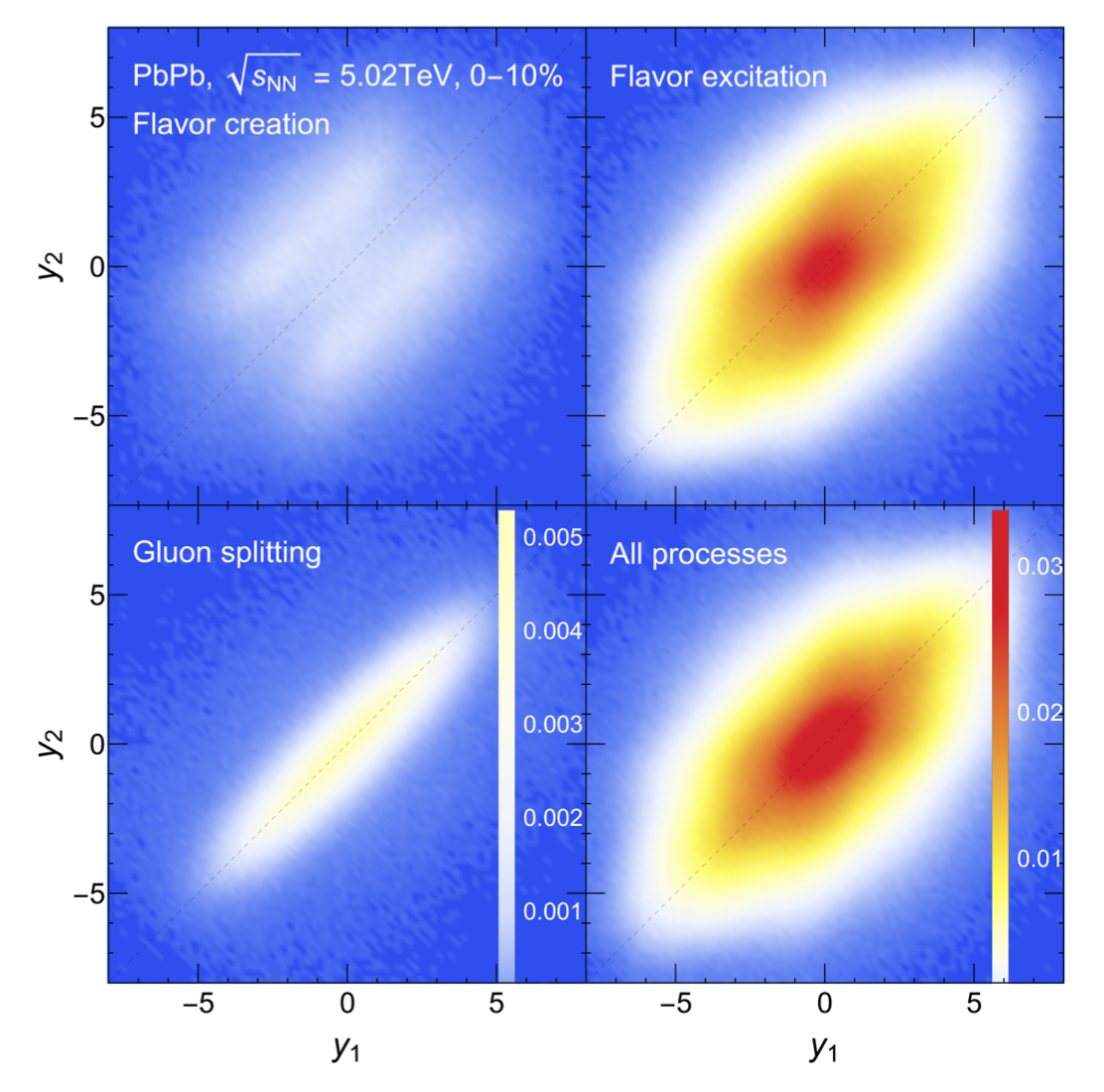}
\caption{The correlation between  the rapidity $y$ of $c$ and $\bar c$ at creation  separated for the  different production processes in Pb-Pb collisions at $\sqrt{s_{\rm NN}}=5.02$  TeV. }
\label{fig.y1y2PbPb}
\end{figure}
\begin{figure}[!htb]
\includegraphics[width=0.4\textwidth]{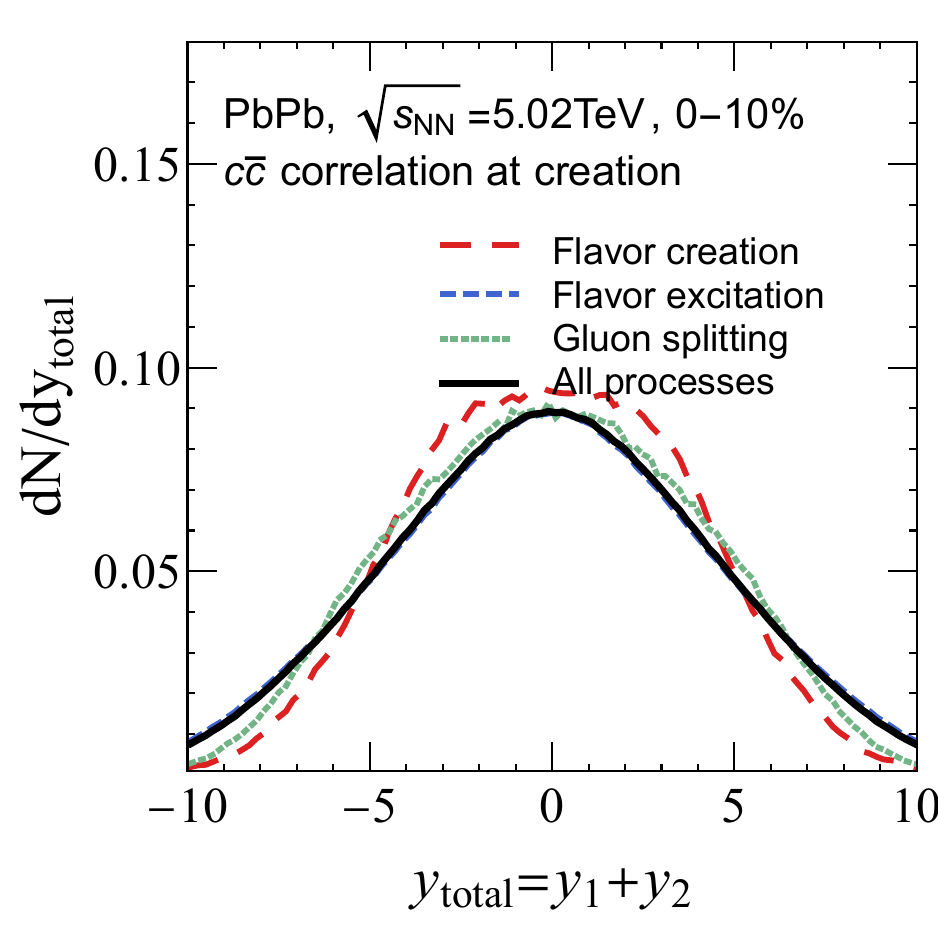}
\caption{The distribution of the total rapidity of $c$ and $\bar c$ for the different production processes in 0-10\% Pb-Pb collisions at $\sqrt{s_{\rm NN}}=5.02$ TeV. All curves are normalized to 1.}
\label{fig.ytotalPbPb}
\end{figure}

As discussed in section \ref{sec.2}, in collisions among hadrons a heavy flavour quark-antiquark ($Q\bar Q$) pair may be produced in three different ways, shown in Fig.~\ref{fig.hardprocess1}. The correlation between $Q$ and $\bar Q$ is rather different for the three different pQCD creation processes.

\subsection{Correlations in rapidity}
The rapidity correlation between $Q$ and $\bar Q$ for the three production processes as well as for all three processes together is displayed in Fig.~\ref{fig.y1y2PbPb}.

We see that the correlations for flavour excitation and gluon splitting are centered around $y_1=y_2$.  For flavor excitation the correlations are rather weak and therefore the $Q$ and $\bar Q$, created at the same vertex, may have rather different rapidities. This means that 
in experiments a large rapidity interval is necessary to register both heavy quarks.
For gluon splitting the distribution is considerably narrower because there is no hard interaction which may cause a strong rapidity shift. For the flavour creating process, however, the distribution has a maximum at a finite rapidity difference between the heavy quarks. Parton distribution functions increase strongly with decreasing $x = p_{parton}/p_{proton}$ values. Therefore collisions with different $x$ values for projectile and target are more probable than symmetric collisions. This is also true for collisions with a sufficient $\sqrt{s}$
value to produce a $Q\bar Q$ pair.

In Fig.~\ref{fig.ytotalPbPb} we display the distributions of the total rapidity, $y_1+y_2$, of the $Q\bar Q$ pairs. They are normalized to one. All peak at $y_{total}=0$  but the distribution caused by the flavor creation is much broader.

\subsection{Correlation in transverse momentum}
\begin{figure}[!htb]
\includegraphics[width=0.45\textwidth]{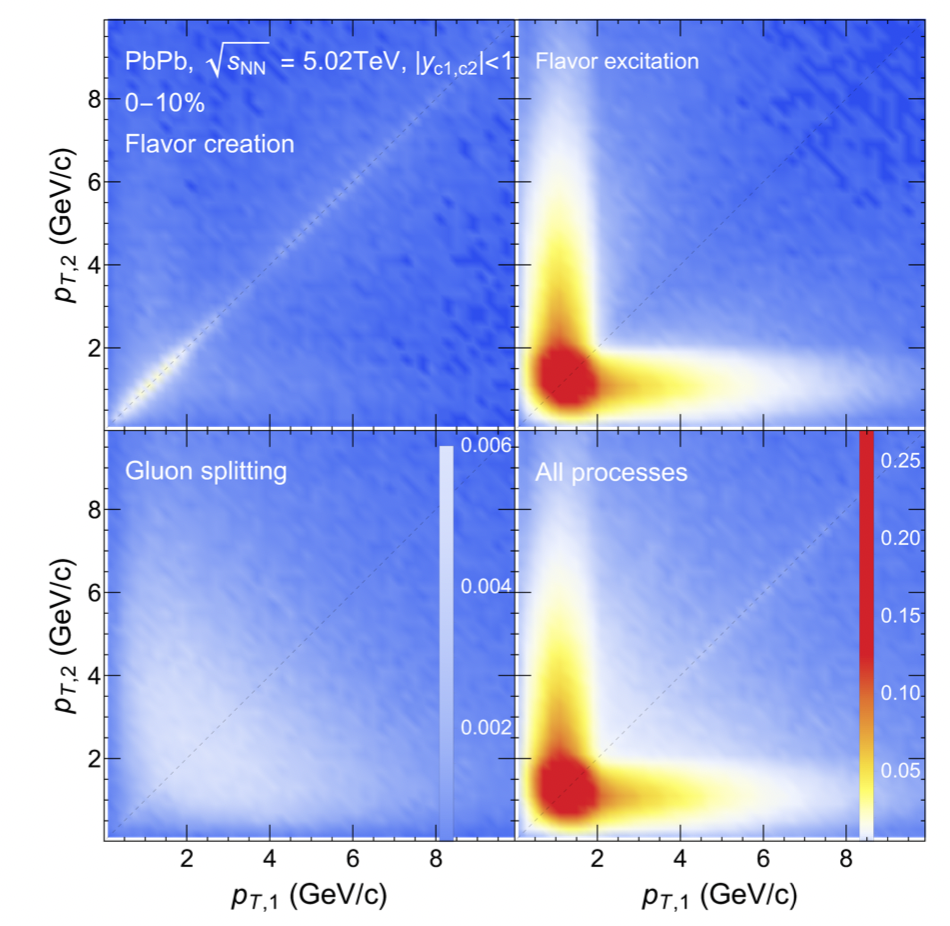}\\
\includegraphics[width=0.45\textwidth]{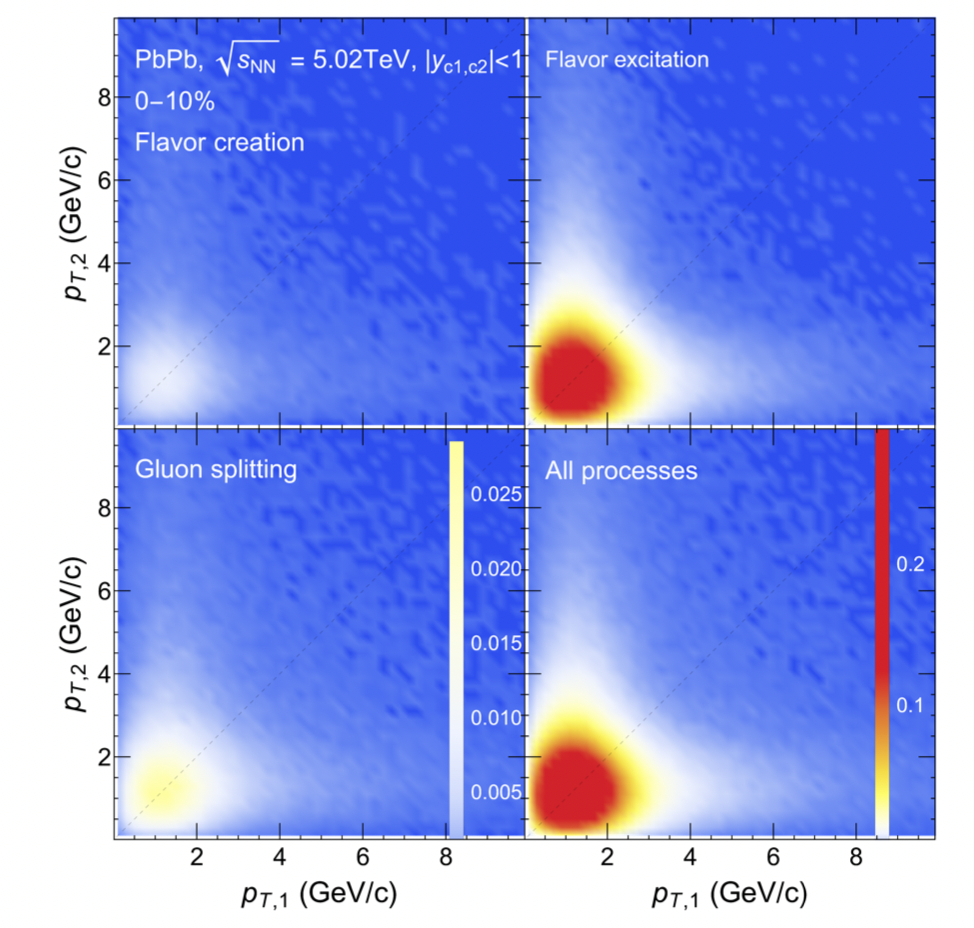}
\caption{The correlation between $p_T$ of $c$ and $\bar c$ at creation (top) and before hadronization (bottom) of different processes in Pb-Pb collisions at $\sqrt{s_{\rm NN}}=5.02$ TeV and at central rapidity $|y|<1$. }
\label{fig.pt1pt2PbPb}
\end{figure}

The correlation between the absolute value of the transverse momentum of the $Q$ and $\bar Q$ of those pairs, where both quarks have a rapidity $|y|<1$, is displayed for central Pb-Pb collisions in Fig.~\ref{fig.pt1pt2PbPb}, as well separated for the three different processes. We display the initial  distribution (top) and the distribution before hadronization, after the passage through the QGP (bottom).  

Due to momentum conservation we expect that initially quarks, coming from the flavour creation process, have very similar transverse momenta and this is indeed seen in this LO process where the $c\bar{c}$ pair is emitted nearly back-to-back.
For flavour excitation, where one of the heavy quarks gets a kick due to the hard interaction, we expect a transverse momentum difference between the heavy quarks, what is also observed.  
\begin{figure}[!htb]
\includegraphics[width=0.5\textwidth]{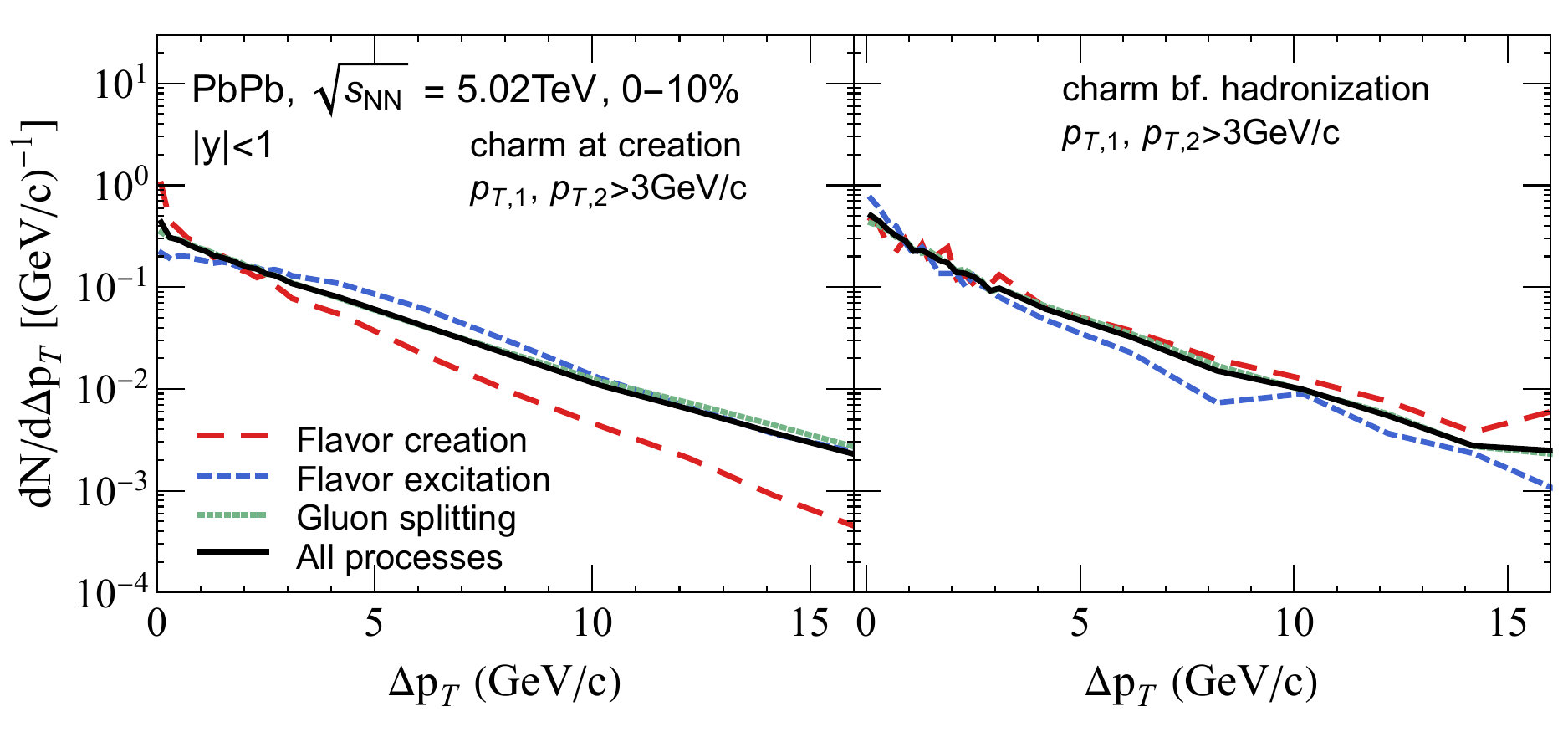}
\caption{The distribution of the transverse momentum difference $|\Delta p_T|$ of  charm quarks with $p_T>3~ \rm GeV/c$ and from different processes in 0-10\% Pb-Pb collisions at $\sqrt{s_{\rm NN}}=5.02$ TeV. All curves are normalized to 1.On the left side we display the distribution at the moment of the production of the $Q\bar Q$ pair, on the right side when the heavy quarks hadronize.}
\label{fig.deltapt}
\end{figure}

Gluon splitting shows a rather weak correlation, which peaks at $p_{T,1} \approx p_{T,2}$.  The dominating process for the creation of a $Q \bar Q$ pair is flavour excitation.

If we compare the top and the bottom figure we see that the correlations are considerably weakened during the travel of the heavy quarks through the QGP, due to their elastic and inelastic collisions.  In addition, the transverse momentum is lowered. The  $p_{T,1} \approx p_{T,2}$ correlation of the pairs created by flavor creation has disappeared and for all creation processes most of the pairs have quite different  $p_{T,1}$ and $ p_{T,2}$ values.  If we investigate smaller systems or more peripheral reactions we expect correlations intermediate between the top and the bottom figure.

This figure shows as well that, if we demand that the $p_T$ of both heavy quarks is larger than 4 GeV, we select pairs which are created by flavour excitation and not by flavour creation. 

To quantify the influence of the QGP on the relative momentum between the heavy quarks
$| \Delta p_T| $ = $|p_{T1}-p_{T2}|$  we display in Fig. \ref{fig.deltapt} this quantity for Pb-Pb collisions at $\sqrt{s_{\rm NN}}=5.02$ TeV for the different production mechanisms, on the left 
hand side at production, on the right hand side before hadronization.  Here both quarks have to have a $p_T >$ 3 GeV and the distributions are normalized to one. We see that initially $Q\bar Q$ pairs created by  flavour creation have a small momentum difference whereas those produced in flavor excitation have on the average a considerably larger $\Delta p_T$. Those from gluon splitting are in between. After traversing the QGP all distributions are rather similar and even that for flavor excitation increases up to $\Delta p_T\to 0$ (see left hand side).  Therefore a $\Delta p_T$
cut cannot serve as possibility to separate the three production processes in central Pb-Pb collisions.

\subsection{Azimuthal correlations}
\begin{figure*}[!htb]
\includegraphics[width=0.8\textwidth]{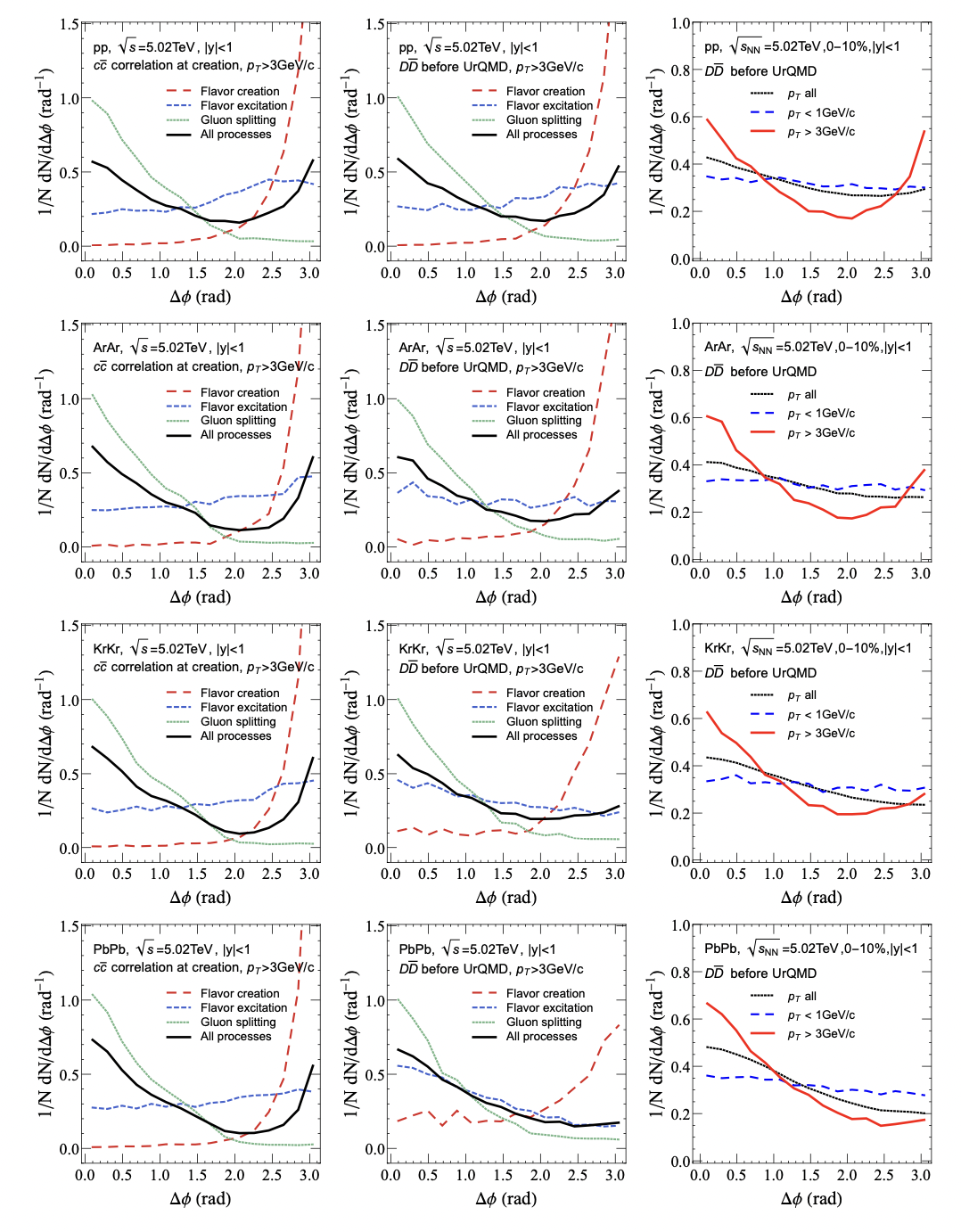}
\caption{Correlations of $D\bar D$ (left) and from different processes (middle and right) in p-p, Ar-Ar, Kr-Kr, and Pb-Pb collisions at $\sqrt{s_{\rm NN}}=5.02$ TeV and central rapidity $|y|<1$. Here the correlations are charm quarks or $D$ mesons from the same vertex. }
\label{fig.azim}
\end{figure*}

\begin{figure*}[!htb]
\includegraphics[width=0.8\textwidth]{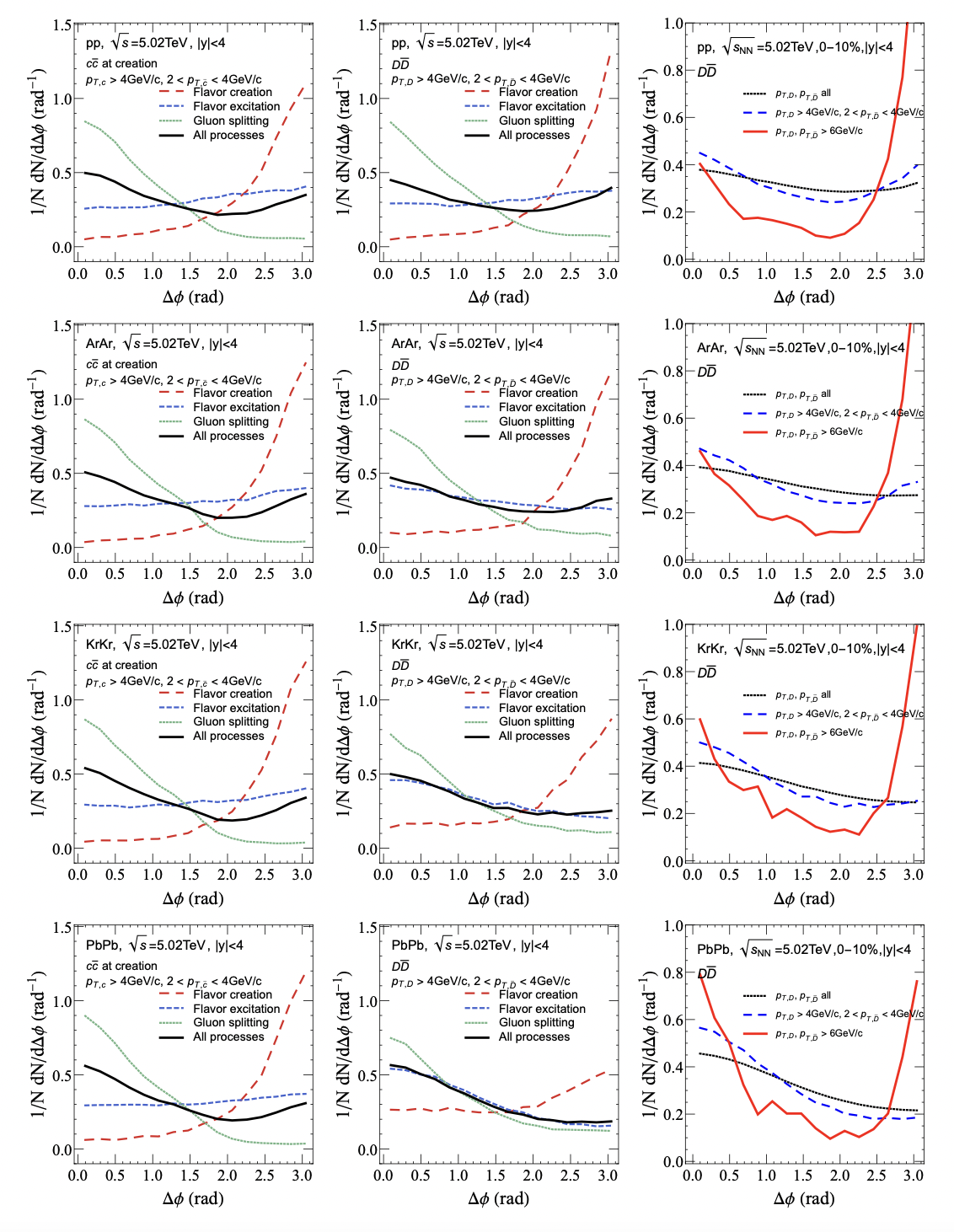}
\caption{Same as Fig. \ref{fig.azim} but for  a wider rapidity range, $|y|<4$.}
\label{fig.azimy4}
\end{figure*}
Next we study the azimuthal correlations between the transverse momenta of the heavy quarks for the different production mechanisms of the $Q\bar Q$ pairs. These are  displayed in Fig.~\ref{fig.azim} for pairs in which both quarks have $|y|<1$. On the left hand side we display the distribution of the difference of the azimuthal angle, $\Delta \phi$, between the $Q$ and $\bar Q$ momenta, separated for the three different production processes, for HQ which both have $p_T>$ 3 GeV.  In the middle column we display this correlation of the HQ after hadronization and in the right column we display the sum of all production processes for
three different $p_T$ cuts for both HQ. From top to bottom we display these correlations
for p-p and central Ar-Ar, Kr-Kr and Pb-Pb collisions.

In the top row, p-p collisions,  we see that each of the production processes creates a different correlation, which is preserved during the hadronization process.   $Q\bar Q$s created by flavour creation show an enhancement for back to back production, as expected by the two body kinematics, which is at the origin of this type of production. Gluon splitting, on the contrary, produces an enhancement around $\Delta \phi = 0$ because the direction of the momentum of both quarks is close to that of the emitting gluon. The correlation between the transverse momenta of those pairs, which are produced by flavour excitation, is weak due to the kick, which one of the heavy quarks obtains  in the hard process. 
Summing up the different correlations the correlation function looks smoother. For $p_T<1$ GeV there are practically no strong azimuthal correlations, which show up when we demand $p_T>3$ GeV. There we see a strong enhancement at $\Delta \phi =0$ and  $\Delta \phi = \pi$ coming from flavour creation and gluon splitting.  We can therefore conclude that the experimentally observed forward-backward enhancement of the correlation function is due to the three different production processes \cite{Zhao:2023ucp}.

When we increase the system size, displayed in the lower rows of Fig.~\ref{fig.azim} for Ar-Ar, Kr-Kr  and Pb-Pb, we observe that the initial correlations are rather similar. The passage through the QGP changes the correlations in a different way. The correlation due to gluon splitting, where both HQ have a similar $p_T$, passes the same region of the QGP and remains rather unchanged whereas that due to flavor creation is strongly weakened. There are two reasons.  One is the collective radial velocity of the QGP, which is communicated by collisions to the heavy quarks. If heavy quarks are emitted back to back, the one with a large momentum component anti-parallel to the radial collective velocity gets decelerated due to collisions and can inverse the direction of its momentum, what is not the case for the other quark which has a large momentum component parallel to the collective velocity.  The other reason is the path length difference in the QGP  of  back to back emitted particles . Both processes weaken the azimuthal correlations between the heavy quarks and depend on the size of the collective expansion velocity (which increases with the system size) as well as on the size of the QGP.  As a consequence, the correlation function for heavy systems at  $\Delta \phi =0$ is enhanced and the peak at $\Delta \phi =\pi$ disappears.  A similar evolution of the correlation function, especially for flavour creation, has been observed in PHSD calculations for Au-Au collisions at $\sqrt{s_{\rm NN}}$ = 200 GeV, using initial $c\bar c$ correlations provided by a tuned Pythia~\cite{Song:2016rzw} calculation,  as well as in other transport approaches~\cite{Zhu:2006er,Nahrgang:2013saa}.
\begin{figure}[!htb]
\includegraphics[width=0.25\textwidth]{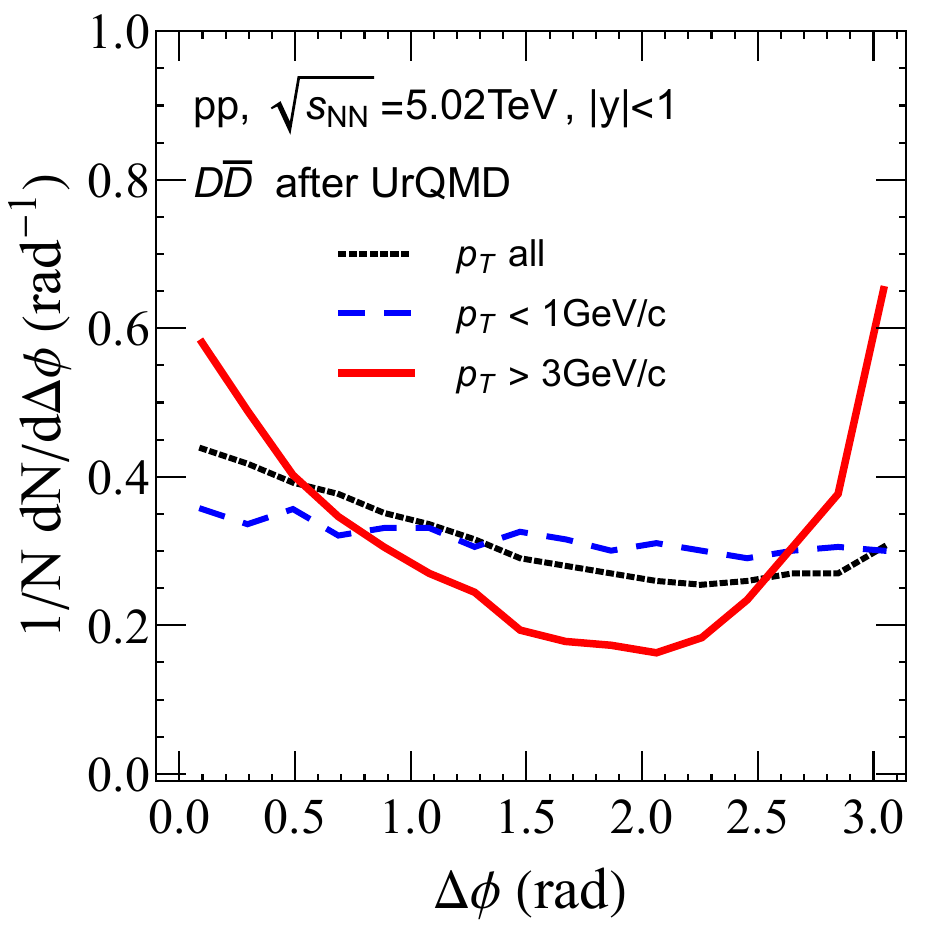}\includegraphics[width=0.25\textwidth]{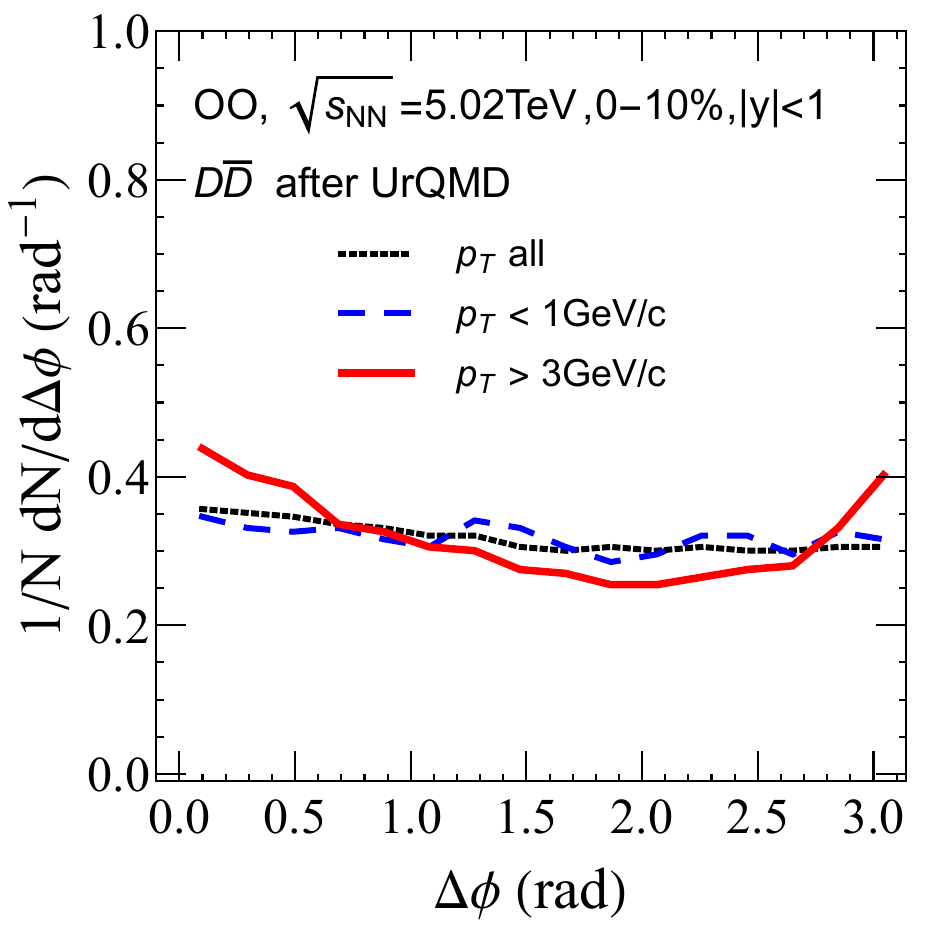}\\
\includegraphics[width=0.25\textwidth]{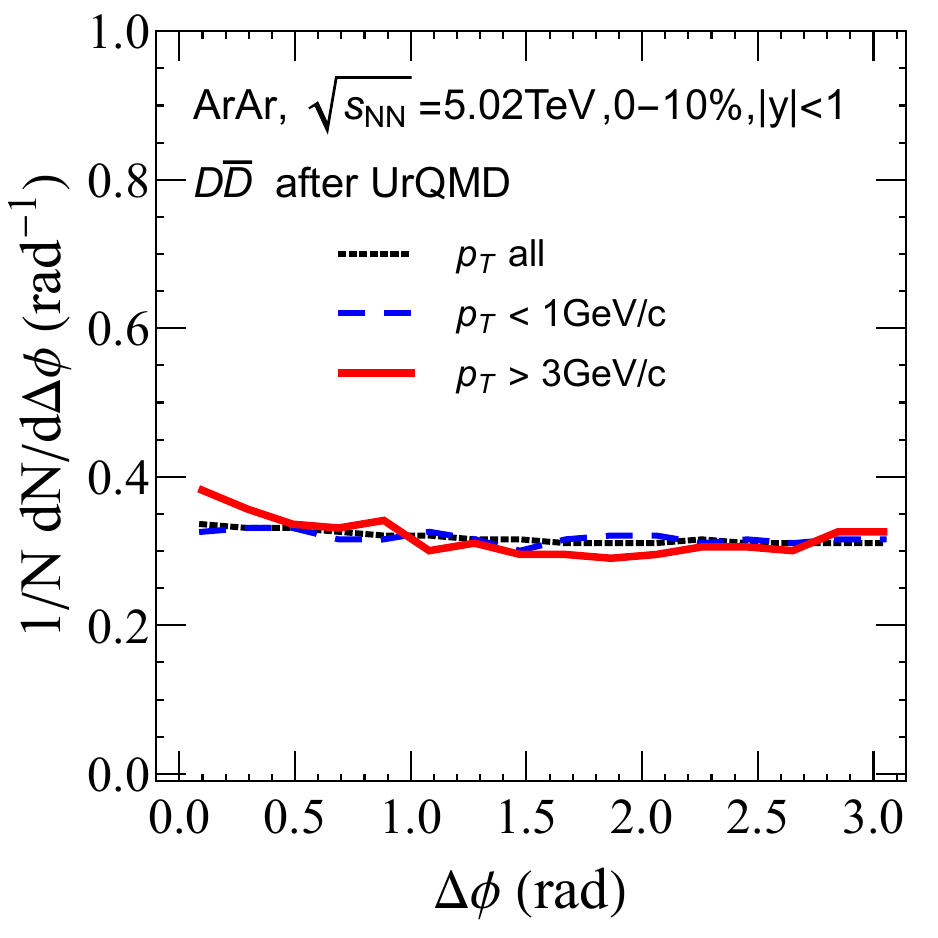}\includegraphics[width=0.25\textwidth]{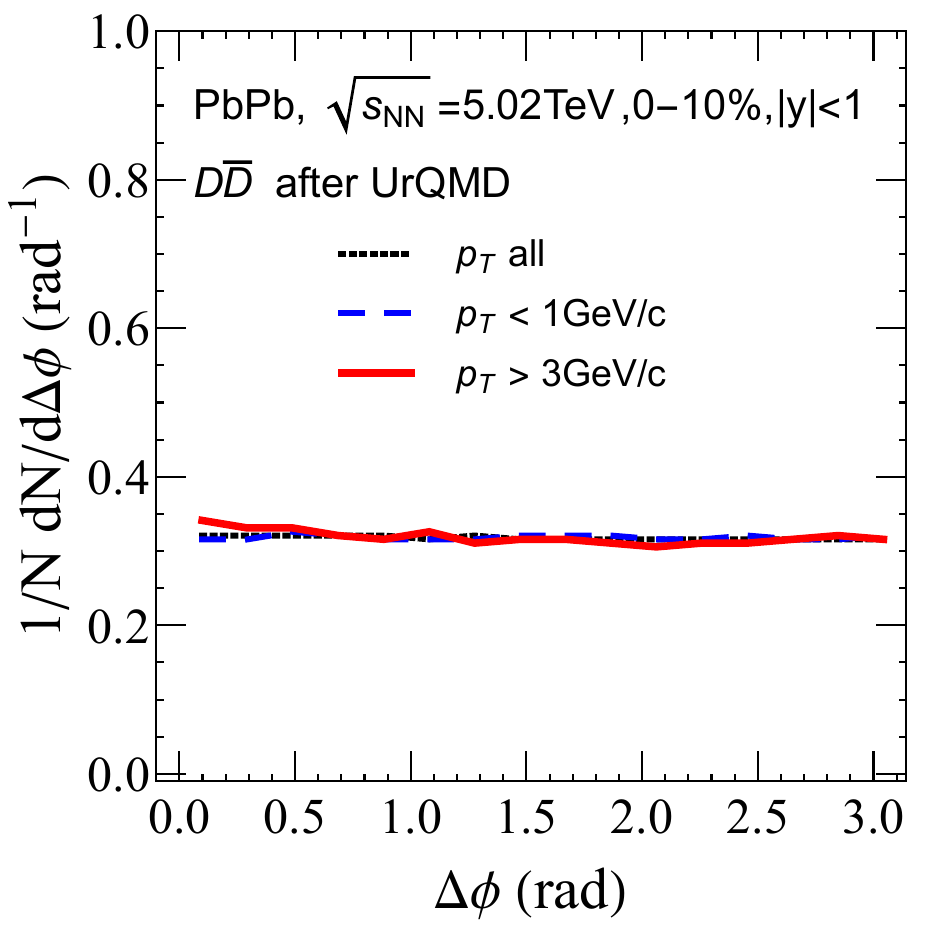}\\
\caption{Correlations of $D\bar D$ in p-p, O-O Ar-Ar, and Pb-Pb collisions at $\sqrt{s_{\rm NN}}=5.02$ TeV and central rapidity $|y|<1$.}
\label{fig.diffver}
\end{figure}

Because it is in discussion to extend the rapidity range of the ALICE detector at LHC, in which heavy quarks can be detected, we display in Fig.~\ref{fig.azimy4} the results if both HQ can be observed in a rapidity interval  of $|y|<4$. We see that the correlations 
are only slightly modified. The flavor excitation process remains rather flat and the enhancement around $\Delta \phi \approx 0$ is reduced for the gluon splitting process.  In the right column we see that for this larger rapidity interval the forward-backward enhancement, seen in p-p, is recovered for Pb-Pb collisions if the $p_T$ of both HQ is larger than 6 GeV. 

\begin{table*}
	\renewcommand\arraystretch{1.5}
	\setlength{\tabcolsep}{1.0mm}
	\begin{tabular}{c|c|c|c}
		\toprule[1pt]\toprule[1pt]
        \multicolumn{1}{c|}{$\Delta\phi>5\pi/6$}&
	\multicolumn{1}{c|}{\rm Flavor creation}& \multicolumn{1}{c|}{\rm Flavor excitation} &    \multicolumn{1}{c}{\rm Gluon splitting} 
        \tabularnewline
		\midrule[1pt]
	$p_{T,D}, p_{T,{\bar D}}$ all & 13.6\% & 81.3\% & 5.1\%   \tabularnewline
 	$p_{T,D}>4, 2<p_{T,{\bar D}}<4$ GeV & 38.3\% & 54.6\% & 7.1\%  
        \tabularnewline
         $p_{T,D}, p_{T,{\bar D}}>6$ GeV & 64.3\% & 30.6\% & 5.1\%  
        \tabularnewline
  		\midrule[1pt]
            \multicolumn{1}{c|}{$\Delta\phi<\pi/6$}&
	\multicolumn{1}{c|}{\rm Flavor creation}& \multicolumn{1}{c|}{\rm Flavor excitation} &    \multicolumn{1}{c}{\rm Gluon splitting} 
        \tabularnewline
        \midrule[1pt]
	$p_{T,D}, p_{T,{\bar D}}$ all & 1.4\% & 70.8\% & 27.8\%   \tabularnewline
 	$p_{T,D}>4, 2<p_{T,{\bar D}}<4$ GeV & 1.9\% & 35.9\% & 62.2\%  
        \tabularnewline
         $p_{T,D}, p_{T,{\bar D}}>6$ GeV & 1.9\% & 23.7\% & 74.4\%  
        \tabularnewline
		\bottomrule[1pt]
	\end{tabular}
	\caption{The fractions of $D\bar D$ are from different processes when did the selection, $\Delta\phi>5\pi/6$ and $\Delta\phi<\pi/6$, in p-p collisions with $\sqrt{s_{\rm NN}}=5.02$ TeV.}
	\label{table1}
\end{table*}
\begin{figure*}[!htb]
\includegraphics[width=1\textwidth]{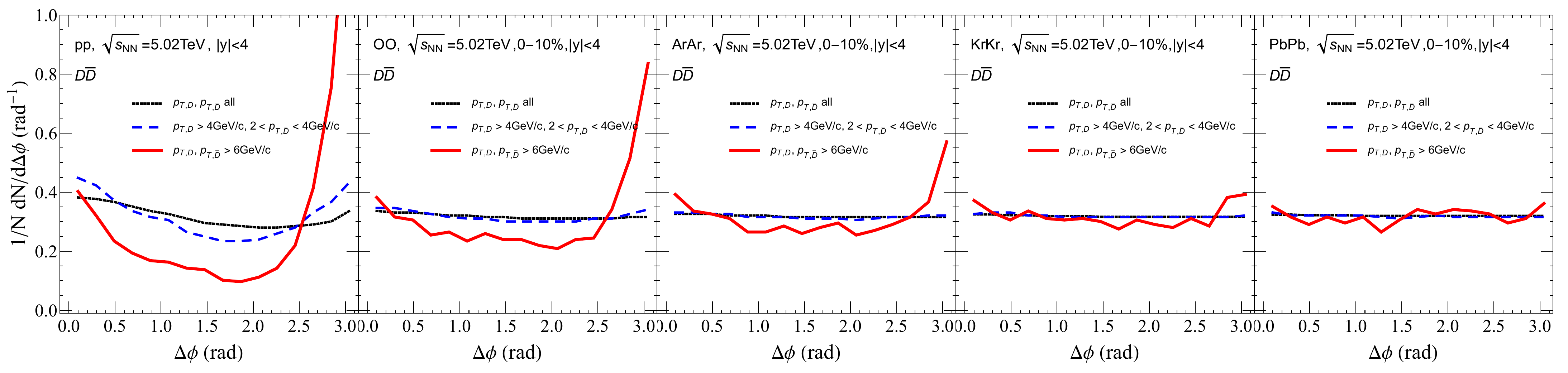}\\
\includegraphics[width=1\textwidth]{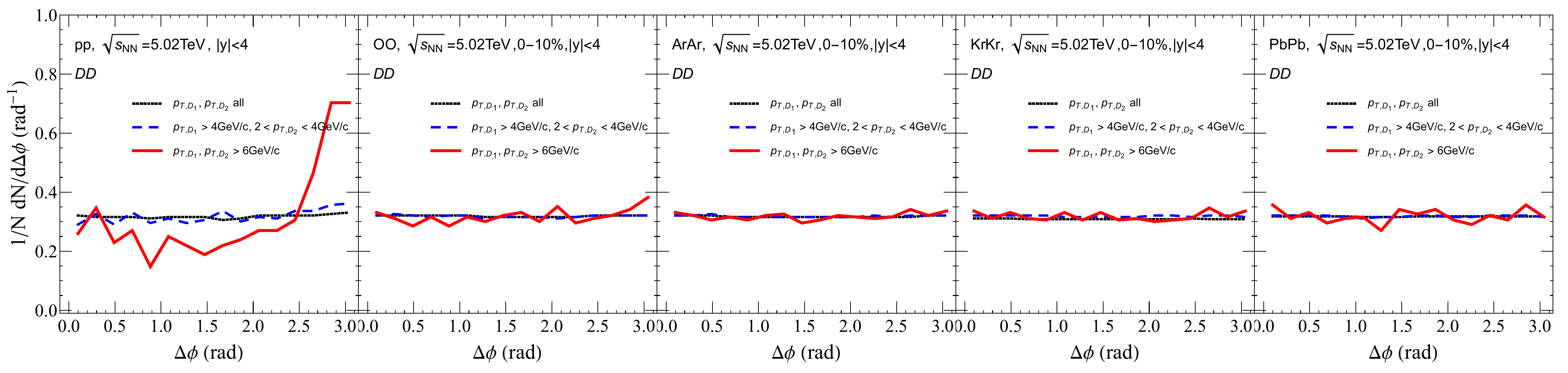}\\
\includegraphics[width=1\textwidth]{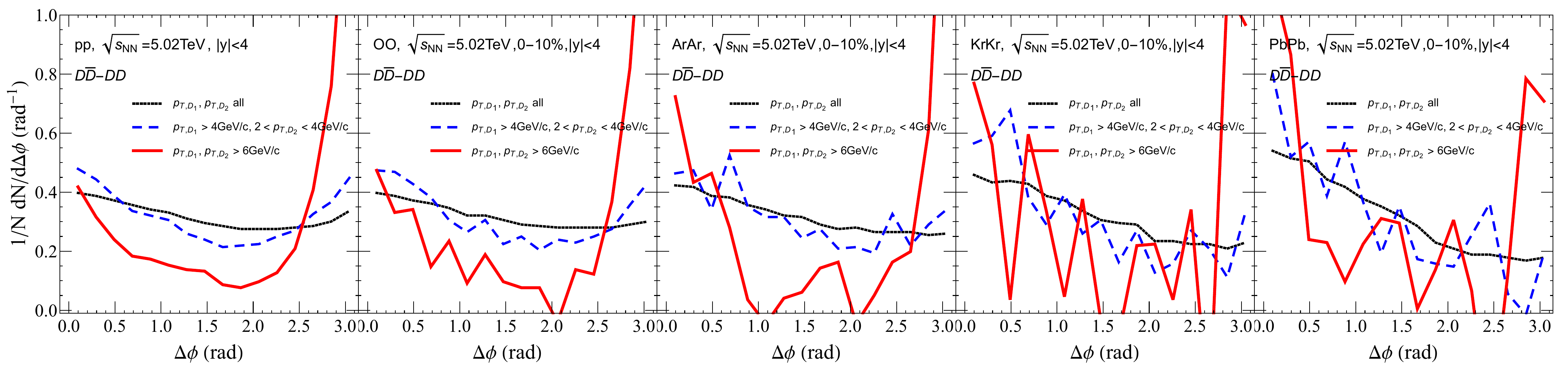}\\
\caption{Correlations of  $D\bar D$ (top), $DD$ (middle), $D\bar D-DD$ (bottom) in p-p, O-O, Ar-Ar, Kr-Kr, and Pb-Pb collisions at $\sqrt{s_{\rm NN}}=5.02$ TeV and central rapidity $|y|<4$. Here $D\bar D-DD$ correlation is counted event-by-event.}
\label{fig.DDall}
\end{figure*}

\begin{figure*}[!htb]
\includegraphics[width=1\textwidth]{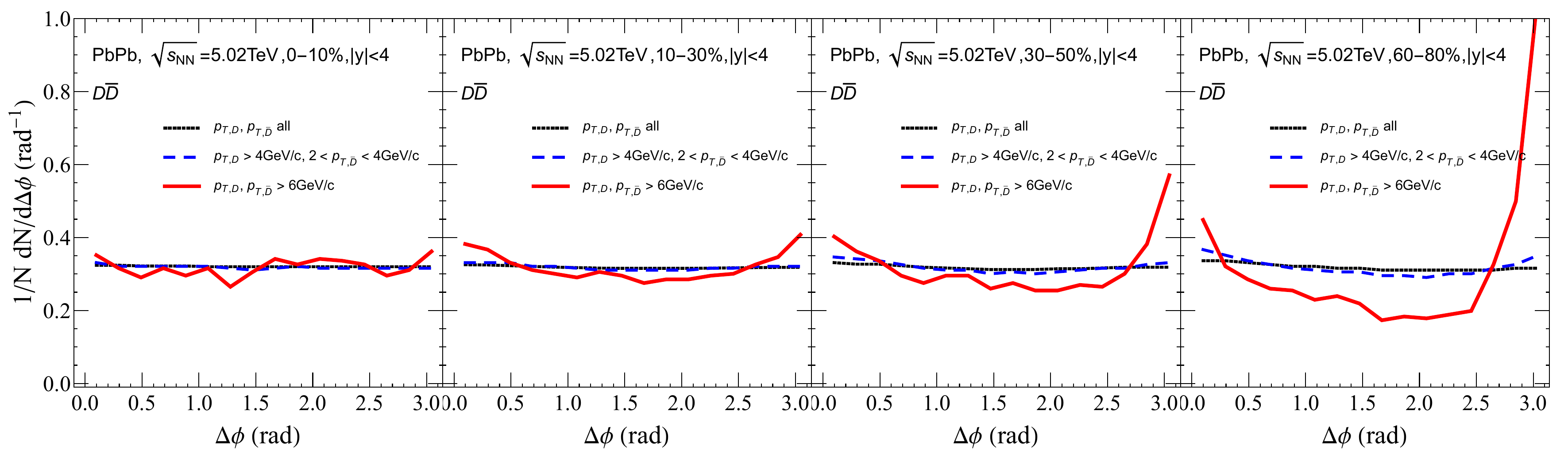}
\caption{Correlations of $D\bar D$ in Pb-Pb collisions at $\sqrt{s_{\rm NN}}=5.02$ TeV and central rapidity $|y|<4$. From left to right are with different centrality bins, [0-10\%], [10-30\%], [30-50\%], and [60-80\%].}
\label{fig.DDbar.PbPb}
\end{figure*}

Up to now we have investigated the ideal situation that one knows at which vertex the $Q\bar Q$ is created. In experiments the vertex cannot be identified and therefore the experimentally observed correlation function is that created by all possible $D\bar D$ pairs. Their number increases with the system size and the $D\bar D$ pairs from different vertices are uncorrelated. The correlation function, obtained by including pairs from different vertices, means the combinatorial background, is shown in Fig.~\ref{fig.diffver} for different system sizes. We see that pairs coming from different vertices, whose number increases strongly as compared to those coming from the same vertex,  wash out rather fast the correlation function. For O-O we see for $p_T >$ 3 GeV for both heavy quarks still a structure of the correlation function but very reduced as compared to p-p. For Ar-Ar the structure of the correlation function has disappeared. This is also the case if one widens the rapidity interval as can be seen from the top row of Fig. \ref{fig.DDall}, where we display the same correlation function. The combinatorial background destroys also here the structure of the correlation function of pairs coming from the same vertex.

The middle row show the correlation function of two $D$-mesons.  In heavy-ion collisions the two $D$-mesons come in their large majority from different pomerons and therefore from  different production vertices and should not show a structure in the correlation function. However, if both $D$ mesons contain heavy quarks from the same pomeron  they are correlatedin EPOS4  \cite{Werner:2023zvo}. This is the case for p-p collisions. 

To quantify these correlations we display in Table~\ref{table1} the fraction of $D\bar D$ mesons which come from the different processes in the intervals $\Delta\phi>5\pi/6$ and $\Delta\phi<\pi/6$, for p-p collisions with $\sqrt{s_{\rm NN}}=5.02$ TeV. We display this percentage for different cuts in the transverse momentum of the $D$ and $\bar D$. We see that without any cut flavour excitation is dominant in both intervals but cuts in the transverse momentum allows to select the pQCD process. For $\Delta\phi>5\pi/6$ this is the flavour creation and for $\Delta\phi<\pi/6$ gluon splitting. So with with a combination of a high $p_T$ cut and a cut in the opening angle one can exclusively study these pQCD processes individually. 

Can this combinatorial background be removed? for experiments event mixing is a possibility, which is however too costly for transport approaches. Having the complete information about all $Q$ and $\bar Q$ quarks available a simpler method is to subtract from the $D\bar D$ correlation the $DD$ correlation, for which the number of possible pairs is almost identical. In the bottom row of Fig.~\ref{fig.DDall} we display the difference between the $D\bar D$ and $DD$ correlation and indeed we recover a structure, which resembles the structure of the original $D\bar D$ correlation function of $D\bar D$ pairs, which are coming from the same vertex, shown in Fig. \ref{fig.azimy4}. It remains to be seen how this recovery potential depends on the acceptance of the detectors.

In heavy ion collisions the structure of the correlation function depends only slightly on the centrality of the collision. This is shown in Fig. \ref{fig.DDbar.PbPb}, where we display for three centrality intervals [0-10\%], [30-50\%] and [60-80\%]  the $D\bar D$ correlation function for three different $p_T$ cuts. In the [30-50\%] centrality bin a slight enhancement at $\Delta \phi \approx \pi$ appears, which becomes even more visible in peripheral collisions, [60-80\%] centrality, where also a small enhancement   at $\Delta \phi \approx 0 $ appears. The structure remains, however less pronounced than in the combinatorial background corrected central collisions (see Fig. \ref{fig.DDall}.)

\section{Conclusion}
\label{sec.7}
We presented a study about the system size dependence of several heavy hadron observables, which have been identified as being sensitive to the interaction between heavy quarks and the QGP, created in heavy ion reactions. These observables have the potential to elucidate further the properties of the expanding QGP, the key objective of the experimental study of ultrarelativistic heavy ion collisions. For this investigation we used the recently advanced EPOS4HQ approach, which has been able to describe the heavy meson single particle spectra in p-p as well as in heavy ion collisions and reproduced in addition quite nicely the presently available data on heavy meson correlations. 

We observe that indeed the interaction of the heavy quarks with the QGP hadrons changes the heavy quark momentum distribution or, the other way round, that the heavy quark spectra carry information about the constituents of the QGP and the interaction strength.  The momentum change depends strongly on the system size. The other source
of the momentum difference between the initial $c$-quark and the finally observed hadron
is hadronization, which is a process which depends only mildly on the system size. A systematic study of the momentum change (or the change of the form of the $p_T$ spectrum) may separate the two sources.

The heavy baryon enhancement at low $p_T$ is a smooth function of the system size and continuous from p-p to Pb-Pb. If verified experimentally, this strengthen the conjecture that this enhancement is a consequence of the production of a QGP, even in a system as small as p-p. 

Azimuthal correlations between the $Q$ and the $\bar Q$ at creation are only weakly modified by shadowing effect. The form of the correlation function is determined by the three different creation mechanisms of $Q\bar Q$ pairs.  Depending on the creation mechanism the momentum difference between the pair particles is rather differently modified when the heavy quarks pass the QGP.  Also hadronization modifies the different $Q\bar Q$ correlation functions.  If one sums over all different production mechanisms the correlation function is much smoother than that for the individual processes. 

In p-p, a cut in $p_T$ and in the opening angle allows to separate experimentally gluon splitting and flavour creation from the other $Q\bar Q$ production mechanisms. This is a promising perspective to study these pQCD processes separately.  

The heavy hadrons coming from the same vertex are still correlated, even in Pb-Pb collisions.  Experimentally the vertex, at which the $Q\bar Q$ pair is produced, cannot be identified.  With increasing system size this leads to a the strong increase of the  multiplicity of heavy quark pairs from different vertices (as compared to those from the same vertex).  This combinatorial background  makes the correlation function featureless and insensitive to the interaction of the heavy quark with QGP partons.

The correlation function can be recovered if one subtracts from the correlation function for $Q \bar Q$ pairs that for  $QQ$ pairs. If this is experimentally possible the system size dependence of the correlation function
is another observable which sheds light on the interaction of heavy quarks with the QGP.

\vspace{1cm}
\noindent {\bf Acknowledgement}: 
This work is supported by the European Union's Horizon 2020 research and innovation program under grant agreement No 824093 (STRONG-2020).

\bibliographystyle{apsrev4-1.bst}
\bibliography{main}

\end{document}